\documentclass[osajnl2,preprint,showpacs,superscriptaddress,12pt]{revtex4} %% use 12pt for preprint option
\usepackage{amsmath,amssymb,graphicx}
\begin{document}
\title{Effects of nonlocal plasmons in gapped graphene micro-ribbon array and 2DEG
on near-field electromagnetic response in the deep-subwavelength regime}

\author{Danhong Huang}

\affiliation{Air Force Research Laboratory, Space Vehicles
Directorate, Kirtland Air Force Base, NM 87117, USA}

\author{Godfrey Gumbs}
\author{Oleksiy Roslyak}

\affiliation{Department of Physics and Astronomy, Hunter College of the
City University of New York, 695 Park Avenue, New York, NY 10065, USA}

\begin{abstract}
A self-consistent theory involving Maxwell equations and a
density-matrix linear-response theory is solved for an
electromagnetically-coupled doped graphene micro-ribbon array and a
quantum-well electron gas sitting at an interface between a
half-space of air and another half-space of a doped semiconductor
substrate which supports a surface-plasmon mode in our system. The
coupling between a spatially-modulated total electromagnetic field
and the electron dynamics in a Dirac-cone of a graphene ribbon, as
well as the coupling of the far-field specular and near-field
higher-order diffraction modes, are included in the derived electron
optical-response function. Full analytical expressions are obtained
with non-locality for the optical-response functions of a
two-dimensional electron gas and a graphene layer with an induced
bandgap, and are employed in our numerical calculations beyond the
long-wavelength limit (Drude model). Both the near-field transmissivity and reflectivity
spectra, as well as their dependence on different configurations of
our system and on the array period, ribbon width, graphene chemical
potential of quantum-well electron gas and bandgap
in graphene, are studied. Moreover, the transmitted E-field
intensity distribution is calculated to demonstrate its connection
to the mixing of specular and diffraction modes of the total
electromagnetic field. An externally-tunable electromagnetic
coupling among the surface, conventional electron-gas and massless
graphene intraband plasmon excitations is discovered and explained.
Furthermore, a comparison is made between the dependence of the
graphene-plasmon energy on the ribbon width and chemical potential
in this paper and the recent experimental observation given by Ju,
{\em et al.\/}, [Nature Nanotechnology, {\bf 6}, 630 (2011)] for a
graphene micro-ribbon array in the terahertz-frequency range.
\end{abstract}

\maketitle

\section{Introduction}
\label{sect1}

There has been a high-level of research and device interests on both the electronic
and optical properties of two-dimensional (2D) graphene
material\,\cite{giem,special,ziegler1,orlita,sarma1} since the report on the first successful
isolation of single graphene layers as well as the related transport
and Raman experiments performed for such layers\,\cite{first}. The
most distinctive difference between a graphene sheet and a
conventional 2D electron gas (EG) in a semiconductor quantum well (QW) is
the electronic band structure, where the energy dispersions of
electrons and holes in the former are linear in momentum space
but they become quadratic for the latter. As a result, electrons and
holes in graphene behave like massless Dirac fermions and show a lot of
unexpected physical phenomena in electron transport and optical
response, including anomalous quantum Hall
effect\,\cite{novoselov,kim}, bare and dressed state Klein
tunneling\,\cite{novoselov1,kim1,roslyak,iurov} and plasmon
excitation\,\cite{wunsch,sarma,oleksiy}, a universal absorption
constant\,\cite{mark,nair}, tunable intraband\,\cite{ju} and
interband\,\cite{basov,wang} optical transitions, broadband $p$--polarization
effect\,\cite{polar}, photoexcited hot-carrier thermalization\,\cite{rana} and transport\,\cite{levitov}, electrically and magnetically
tunable band structure for ballistic transport\,\cite{roslyak2}, field-enhanced mobility in
a doped graphene nanoribbon\,\cite{huang0} and electron-energy loss in
gapped graphene layers\,\cite{gumbs}.
\medskip

It has been found that most of the unusual electronic properties in graphene can be very
well explained by single-particle excitation of electrons. For
diffusion-limited electron transport in doped graphene, the Kubo
linear-response theory\,\cite{kubo}, as well as the Hartree-Fock theory in combination with the self-consistent
Born approximation\,\cite{ahm}, were used. Moreover, the semiclassical
Boltzman theory was applied in both the linear\,\cite{linear} and
nonlinear\,\cite{huang0} response regimes. For the collective
excitation of electrons in graphene, previous studies found it has an
important role in the dynamical screening of the
electron-electron interaction\,\cite{wunsch,sarma,oleksiy,gumbs,plasmon}. However, the
EM response of graphene materials has received
relatively less attention, especially for low-energy intraband
optical transions\,\cite{ju,ziegler,soljacic}.
\medskip

For any charged-particle systems, there always exists an electronic
coupling based on the intrinsic electron-electron interaction in addition to
the system response to an external electromagnetic field or an external electron (ion)
beam. This electronic coupling can be either a classical direct coulomb
interaction between any two charged particles with no requirement
for an overlap of their wave functions, or quantum-mechanical
tunneling and exchange coulomb interaction with a tail overlap of their wave
functions\,\cite{book}. For the same charged-particle systems, there
exists another electromagnetic coupling based on the light-electron
interaction\,\cite{huang1}. This electromagnetic coupling results
from an induced optical-polarization field in the system due to
dipole coupling to an incident light wave. The total electromagnetic
field, including the induced optical-polarization field, should
satisfy the macroscopic Maxwell equations\,\cite{huang2}. However,
the induced optical-polarization field in the system is determined
by the microscopic density-matrix equations\,\cite{huang3}.
Therefore, a self-consistent theory is required for
studying the electromagnetic response of such a charged system to
incident light.
\medskip

Physically, a statistical ensemble average of dipole moments in an electronic system
with respect to all quantum states will give rise to
an induced optical-polarization field\,\cite{offdiagonal}. This
provides a connection between the microscopic density-matrix equations
and the macroscopic Maxwell equations\,\cite{huang3}. The
former describes the full electron dynamics excited
by a total electromagnetic field, including optical pumping, energy
relaxation, optical coherence and its dephasing, and carrier
scattering with impurities, lattice and other charged
carriers\,\cite{scattering,scattering1,scattering2}. When the
incident field is weak, a self-consistent linear-response theory to the total
electromagnetic field (not just the incident light) can be applied
to an electronic system as a leading-order approximation.
\medskip

Experimentally, the inelastic light scattering\,\cite{platzman} and
electron-energy-loss spectroscopy\,\cite{review} were widely used
for probing the bulk and surface electronic coupling, respectively,
in a charged-particle system. On the other hand, the elastic light
scattering\,\cite{aam} and the optical absorption spectroscopy\,\cite{absorption} were extensively applied for studying the surface
and thin-film electromagnetic coupling in either a conductive or a
dielectric system. For a smooth conductive surface, only
electronic excitations in the long-wavelength limit can be coupled to an
incident light wave, where a Drude-type model is usually adopted in
spectroscopy analysis. For a patterned conductive surface in the deep sub-wavelength regime, on the
other hand, surface light scattering can be used to explore surface electronic excitations in the
system beyond the long-wavelength limit, where a wavenumber-dependent (nonlocal)
optical response of electrons, which is crucial for the near-field distribution and the Landau damping of plasmons\,\cite{book}, must be employed to analyze the light
reflectivity and transmissivity spectra.
\medskip

In this paper, we consider a model in which a two-dimensional
electron gas (2DEG) and a doped linear graphene micro-ribbon array
(electronically isolated from each other) are put on the surface of
a doped semi-infinite semiconductor substrate. Therefore, we expect
to see the excitations of the localized surface plasmon and the
conventional 2DEG plasmon, as well as the 2D Dirac-fermion plasmon
excitation, in such a system in the presence of an incident
electromagnetic wave. More interestingly, we expect there to exist
strong electromagnetic couplings among these three collective
excitations in the deep sub-wavelength regime, i.e. with an array
period much less than the incident wavelength, and these couplings
can be externally tuned by the doping densities in a graphene layer
or in a QW, as well as by the driven 2DEG in the QW or by the
induced bandgap in the graphene.
The existence of an bandgap in the graphene not only modifies the Landau damping of plasmons\,\cite{oleksiy},
but also introduces the interplay between Dirac-like and Schr\"odinger-like electrons in the electromagnetic response.
To demonstrate the predicted strong
and tunable electromagnetic coupling among three plasmon excitations
in our model system, numerical calculations and their comparisons
are presented for both the transmissivity and reflectivity spectra
with different configurations of our model system and various array
periods, ribbon widths, chemical potentials of graphene, and induced bandgaps in graphene.
\medskip

The rest of the paper is organized as follows. In Sec.\,\ref{sect2},
we derive the self-consistent Maxwell equations and the
density-matrix linear-response theory within the random-phase
approximation for our model system which consists of a combined 2DEG
in an InAs QW and a linear graphene micro-ribbon array doped by
electrons on the surface of a semi-infinite $n$-doped GaAs
substrate. This self-consistent theory, along with the detailed
expressions for a linear matrix equation in Appendix\ A, is then
applied to study the transmissivity and reflectivity spectra of an
incident plane wave for various structural-geometry and material
parameters. In Sec.\,\ref{sect3}, numerical results, including the
$p$--polarized far- and near-field transmissivity spectra as well as
the transmitted $p$--polarized E-field intensity distribution, are
presented to demonstrate and explain the existence of an
externally-tunable electromagnetic coupling among the localized
surface plasmon, 2DEG plasmon and massless graphene plasmon
excitations in our system. The conclusions drawn from these results
are briefly summarized in Sec.\,\ref{sect4}.

\section{Model and Theory}
\label{sect2}

\begin{figure}[htbp]
%\centerline{\epsfig{file=figure1.eps,width=4.0in}}
\centerline{\includegraphics[width=4.0in]{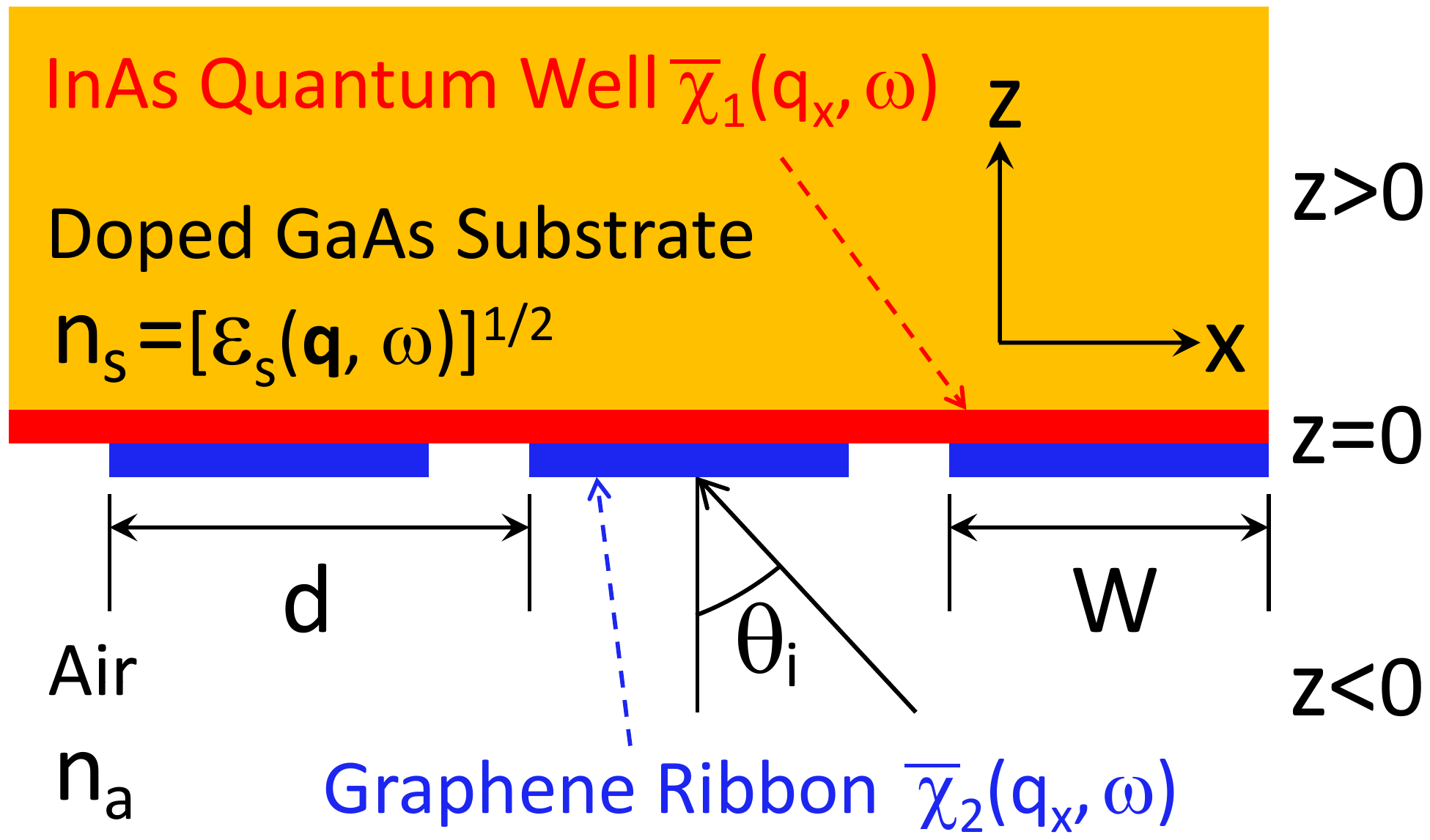}}
\caption{(Color online) Schematic representation of a graphene
micro-ribbon array (blue) with period $d$ and ribbon width ${\cal
W}$ and an InAs quantum well (red). The lower half-space ($z<0$) is
filled with air with refractive index $n_a=1$, whereas the upper
half-space ($z>0$) is filled with a doped semi-infinite GaAs bulk
having a complex dielectric function $\epsilon_s(q,\,\omega)$ or a
complex refractive index $n_s=\sqrt{\epsilon_s(q,\,\omega)}$. Both
the graphene micro-ribbon array [with an optical-response function
$\bar{\chi}_2(q_x,\,\omega)$] and the InAs quantum well [with an
optical-response function $\bar{\chi}_1(q_x,\,\omega)$] sit on the
surface ($z=0$) of the semi-infinite GaAs bulk. A plane-wave
electromagnetic field is incident from the $z<0$ side with an
incident angle $\theta_i$.} \label{f1}
\end{figure}

It is  shown in Fig.\,\ref{f1} that the half-space $z<0$ is air with
the refractive index $n_a=1$ whereas the half-space $z>0$ is
occupied by  a doped semi-infinite GaAs substrate with a dynamical
dielectric function $\epsilon_s(q,\,\omega)$ which supports a
localized surface-plasmon (SP) mode. At the interface $z=0$, there
are a periodic electron-doped graphene micro-ribbon array (GMRA) and
an InAs quantum well (QW) for a 2DEG.
Optically, the GMRA and QW can be considered to be in one plane.
The reason for this is that the sheet thickness is much smaller
than the wavelength and the decay length of the impinging
light. They can, however, still be electrically separated from
each other by an energy barrier\,\cite{huang1,huang2}.
A plane-wave electromagnetic (EM) field incident from $z<0$
is diffracted by the GMRA into both the reflection ($z<0$) and
transmission ($z>0$) regions. The diffraction modes of the EM field are
modulated by the induced optical polarization at the interface with
the GMRA, which will be electromagnetically coupled to SP and 2DEG
plasmons.
\medskip

We assume that the GMRA is periodic in the $x$ direction, and seek
the solutions based on a plane-wave expansion for a single
frequency. Setting $q_y=0$ for simplicity and denoting the
transverse electric component of the EM field below the interface by ${\bf
E}_<$ and above by ${\bf E}_>$, we obtain\,\cite{huang1,huang2,huang3}

\begin{eqnarray}
\left.{\bf E}_<(x,z\right|\omega)&=&e^{iq_xx+i\eta z}\,\left[\begin{array}{c}
\left.A_x(q_x\right|\omega)\\ \left.A_y(q_x\right|\omega)\\
\left.-(q_x/\eta)\,A_x(q_x\right|\omega)
\end{array}\right]
\nonumber\\
&+&\sum\limits_{n=-\infty}^{\infty}\,e^{iq_nx-i\beta_{1,n}z}
\left[\begin{array}{c} \left.B_x( q_n\right|\omega)\\ \left.B_y( q_n\right|\omega)\\
\left.(q_n/\beta_{1,\,n})\,B_x(q_n\right|\omega)
\end{array}\right]\ ,
\ \ \ \ \mbox{for $z<0$}\ ,
\label{e1}
\end{eqnarray}

\begin{eqnarray}
\left.{\bf E}_>(x,z\right|\omega)&=&
\sum\limits_{n=-\infty}^{\infty}\,e^{iq_nx+i\beta_{2,n}z}
\left[\begin{array}{c} \left.C_x( q_n\right|\omega)\\
\left.C_y(q_n\right|\omega)\\
\left.-(q_n/\beta_{2,\,n})\,C_x( q_n\right|\omega)
\end{array}\right]\ ,
\ \ \ \ \mbox{for $z>0$}\ .
\label{e2}
\end{eqnarray}
Here, $n=0,\,\pm 1,\,\pm 2,\,\cdots$ label all reciprocal lattice
vectors $nG\equiv n(2\pi/d)$, $d$ is the period of the GMRA, $q_n=q_x+nG$,
$q_x=(\omega/c)\,n_a\sin\theta_i$,
$\eta=(\omega/c)\,n_a\cos\theta_i$, $n_a$ is the refractive index of
air, $\theta_i$ is the angle of incidence, and $\omega$ the angular
frequency of the incident EM field. In addition, $\beta_{1,n}$ and
$\beta_{2,n}$ in Eqs.\,(\ref{e1}) and (\ref{e2}) are determined by
the following dispersion relations for transverse EM fields

\begin{equation}
\left[
\begin{array}{c}
\left(\beta_{1,n}\right)^2\\ \left(\beta_{2,n}\right)^2
\end{array}\right]=\frac{\omega^2}{c^2}\left[
\begin{array}{c} n_a^2\\ \epsilon_s(q_n,\beta_{2,n},\,\omega)
\end{array}\right]-q_n^2\ ,
\label{e3}
\end{equation}
where $c$ is the speed of light in vacuum,
$\epsilon_s(q_x,\beta,\,\omega)_s$ is the dielectric function of the
doped semiconductor substrate, and we require that
$\mbox{Im}(\beta_{1,n}),\, \mbox{Im}(\beta_{2,n})\geq 0$ so as to
ensure a finite value for the reflected or transmitted EM field at
$z=\pm\infty$.
\medskip

Formally, if we keep only the term with $n=0$ in the summations in
Eqs.\,(\ref{e1}) and (\ref{e2}), the GMRA in our model will simply
change to a graphene sheet. For this situation, after removing the doping
in the GaAs substrate and the 2DEG in the InAs QW, the
formalism in this paper will reduce to the previous ones\,\cite{ziegler,soljacic} for a
transverse EM response from a single graphene
sheet.
\medskip

In the long-wavelength limit, the transverse dielectric function
$\epsilon_s(q_x,\beta,\,\omega)$ introduced in Eq.\,(\ref{e3}) for
the doped semiconductor substrate is found to
be\,\cite{huang1,huang2,huang3}

\begin{equation}
\epsilon_s(q_x,\beta;\omega)=\epsilon_b\left[1-\frac{\Omega^2_{\rm pl}}{\omega(\omega+i\gamma_0)}\right]\ ,
\label{difunc}
\end{equation}
where $\Omega_{\rm pl}=\sqrt{n_{\rm
3D}e^2/\epsilon_0\epsilon_bm^\ast}$ is the bulk plasma frequency,
$n_{\rm 3D}$ is the doping concentration in GaAs, $\hbar\gamma_0$ is the
homogeneous level broadening, $m^\ast$ is the effective mass of
electrons in the substrate and $\epsilon_b$ is the relative
dielectric constant of the semiconductor substrate.
\medskip

Once the electric field ${\bf E}$ is obtained by using the Maxwell
equations, the magnetic field ${\bf H}$ for non-magnetic materials
can be obtained from

\begin{equation}
{\bf H}=\frac{-i}{\omega\mu_0}\,\mbox{\boldmath$\nabla$}\times{\bf E}\ ,
\label{e4}
\end{equation}
where $\mu_0$ is the vacuum permeability. For $q_y=0$ and $z<0$,
Equation\ (\ref{e4}) explicitly leads to

\begin{equation}
\left.H_<^x(x,z\right|\omega)=-\frac{1}{\omega\mu_0}
\left[\eta\,e^{iq_xx+i\eta z}\left.A_y(q_x\right|\omega)
-\sum\limits_{n=-\infty}^{\infty}\,\beta_{1,n}\,
e^{iq_nx-i\beta_{1,n}z}\left.B_y(q_n\right|\omega)\right]\ ,
\label{e5}
\end{equation}

\begin{equation}
\left.H_<^y(x,z\right|\omega)= \omega\epsilon_0n_a^2
\left[\frac{1}{\eta}\, e^{iq_xx+i\eta z}\left.A_x(q_x\right|\omega)
-\sum\limits_{n=-\infty}^{\infty}\,\frac{1}{\beta_{1,n}}\,
e^{iq_nx-i\beta_{1,n}z}\left.B_x(q_n\right|\omega)\right]\ ,
\label{e6}
\end{equation}

\begin{equation}
\left.H_<^z(x,z\right|\omega)= \frac{1}{\omega\mu_0}
\left[q_x\,e^{iq_xx+i\eta z}\left.A_y(q_x\right|\omega)
+\sum\limits_{n=-\infty}^{\infty}\,q_n\,
e^{iq_nx-i\beta_{1,n}z}\left.B_y(q_n\right|\omega)\right]\ .
\label{e7}
\end{equation}
In a similar way, the magnetic field ${\bf H}_>$ in the region of
$z>0$ can also be obtained from

\begin{equation}
\left.H_>^x(x,z\right|\omega)=-\frac{1}{\omega\mu_0}
\sum\limits_{n=-\infty}^{\infty}\,\beta_{2,n}\,
e^{iq_nx+i\beta_{2,n}z}\left.C_y(q_n\right|\omega)\ , \label{e8}
\end{equation}

\begin{equation}
\left.H_>^y(x,z\right|\omega)= \omega\epsilon_0
\sum\limits_{n=-\infty}^{\infty}\,\frac{1}{\beta_{2,n}}\,
e^{iq_nx+i\beta_{2,n}z}\,\epsilon_s(q_n,\beta_{2,n},\,\omega)\left.C_x(
q_n\right|\omega)\ , \label{e9}
\end{equation}

\begin{equation}
\left.H_>^z(x,z\right|\omega)=\frac{1}{\omega\mu_0}
\sum\limits_{n=-\infty}^{\infty}\,q_n\,
e^{iq_nx+i\beta_{2,n}z}\left.C_y(q_n\right|\omega)\ . \label{e10}
\end{equation}
\medskip

Electronically, the GMRA and 2DEG can be separated by an undoped
wide-gap semiconductor barrier layer. Optically, however, the GMRA and
2DEG can both be regarded as two-dimensional sheets since their thickness
are much less than the EM-field wavelength considered. Therefore, the
four independent boundary conditions for transverse EM fields at the
interface $z=0$ can be written as\,\cite{huang1,huang2,huang3}

\begin{eqnarray}
&&\left.E_>^x(x,z=0\right|\omega)-\left.E_<^x(x,z=0\right|\omega)=0\ , \label{e11}\\
&&\left.E_>^y(x,z=0\right|\omega)-\left.E_<^y(x,z=0\right|\omega)=0\ , \label{e12}\\
&&\left.H_>^x(x,z=0\right|\omega)-\left.H_<^x(x,z=0\right|\omega)=\left.-i\omega P_s^y(x\right|\omega)+\left.\alpha_s^y(x\right|\omega)\ , \label{e13}\\
&&\left.H_>^y(x,z=0\right|\omega)-\left.H_<^y(x,z=0\right|\omega)=\left.i\omega P_s^x(x\right|\omega)+\left.\alpha_s^x(x\right|\omega)\ , \label{e14}
\end{eqnarray}
where ${\bf
P}_s\equiv(P_s^x,\,P_s^y)$ is the induced sheet optical-polarization field (related to the polarization current) while
$\mbox{\boldmath$\alpha$}_s\equiv(\alpha_s^x,\,\alpha_s^y)$ is the induced sheet conduction-current
density. Furthermore, we can write $\left.{\bf
P}_s(x\right|\omega)$ as\,\cite{huang2,huang3}

\begin{equation}
\left.{\bf P}_s(x\right|\omega)=\sum\limits_{n=-\infty}^{\infty}\left.{\bf P}_s(q_n\right|\omega)\,e^{iq_nx}\ ,
\label{e15}
\end{equation}
where

\[
\left[\begin{array}{c}
\left.P_s^x(q_n\right|\omega)/\epsilon_0\\ \left.P_s^y(q_n\right|\omega)/\epsilon_0
\end{array}\right]=\chi_1(q_n,\,\omega)\left[\begin{array}{c}
\left.C_x(q_{n}\right|\omega)\\ \left.C_y(q_{n}\right|\omega)
\end{array}\right]
\]
\begin{equation}
+\zeta\sum\limits_{n^\prime=-\infty}^{\infty}\,\chi_2(q_{n^\prime},\,\omega)\,{\rm sinc}[(n^\prime-n)\pi\zeta]
\left[\begin{array}{c}
\left.C_x(q_{n^\prime}\right|\omega)\\ \left.C_y(q_{n^\prime}\right|\omega)
\end{array}\right]\ .
\label{e16}
\end{equation}
In Eq.\,(\ref{e16}), $\zeta={\cal W}/d$, ${\cal W}$ is the width of
a graphene micro-ribbon, ${\rm sinc}(x)\equiv \sin x/x$, and
$\chi_1(q_x,\,\omega)$ and $\chi_2(q_x,\,\omega)$ are the 2DEG and
the graphene-ribbon polarizabilities, respectively. The multi-ribbon
effect represented by the summation over $n^\prime$, as well as the
mode-mixing effect with $n\neq n^\prime$, are both taken into
account by the second term of Eq.\,(\ref{e16}). Moreover, the
coupling between a spatially-dependent total EM field and the electron
dynamics in an energy band, due to the non-locality in an optical
response of electrons, is also included in Eq.\,(\ref{e16}).
\medskip

Similarly, using Ohm's law we can write the sheet conduction-current
density as\,\cite{huang2,huang3}

\begin{equation}
\left.\mbox{\boldmath$\alpha$}_s(x\right|\omega)=\sum\limits_{n=-\infty}^{\infty}\left.\mbox{\boldmath$\alpha$}_s(q_n\right|\omega)\,e^{iq_nx}\ ,
\label{e17}
\end{equation}
where

\[
\left[\begin{array}{c}
\left.\alpha_s^x(q_n\right|\omega)\\ \left.\alpha_s^y(q_n\right|\omega)
\end{array}\right]=\sigma_1(q_{n},\,\omega)\left[\begin{array}{c}
\left.C_x(q_{n}\right|\omega)\\ \left.C_y(q_{n}\right|\omega)
\end{array}\right]
\]
\begin{equation}
+\zeta\sum\limits_{n^\prime=-\infty}^{\infty}\,\sigma_2(q_{n^\prime},\,\omega)\,{\rm sinc}[(n^\prime-n)\pi\zeta]
\left[\begin{array}{c}
\left.C_x(q_{n^\prime}\right|\omega)\\ \left.C_y(q_{n^\prime}\right|\omega)
\end{array}\right]\ ,
\label{e18}
\end{equation}
and $\sigma_1(q_x,\,\omega)$ and $\sigma_2(q_x,\,\omega)$ are the real
2DEG and graphene-ribbon optical conductivities, respectively.
\medskip

For an incoming $p$--polarized EM field, we get $\left.A_x(q_x\right|\omega)=\eta
H_0/(\omega\epsilon_0n_a^2)$ and $\left.A_y(q_x\right|\omega)=0$,
where $H_0$ is the magnetic component and $\epsilon_0$ is the
permittivity of free space.
\medskip

Combining the sheet optical polarization in Eq.\,(\ref{e15}) and the
sheet-current density in Eq.\,(\ref{e17}), we can write the
optical-response function for $j=1,\,2$ in a compact form given by

\begin{equation}
\bar{\chi}_j(q,\,\omega)=\chi_j(q,\,\omega)+\frac{i\sigma_j(q,\,\omega)}{\omega\epsilon_0}\ ,
\label{e19}
\end{equation}
where $q=\sqrt{q_x^2+q_y^2}$.
\medskip

We notice that there exists an anisotropy in the polarizability of a graphene nanoribbon along the longitudinal and transverse directions\,\cite{ribbon}, respectively,
and it can be attributed to a subband quantization due to finite-size effect.
However, this anisotropy can be neglected for the width of ribbons in the micrometer range.
In other words, we adopt in this paper the isotropy of the micro-ribbon array instead of anisotropy of the nanoribbon array.
Within the random-phase approximation, for graphene ribbons in the micrometer range we
calculate the nonlocal ($q\neq 0$) optical-response function
$\bar{\chi}_2(q,\,\omega)$ as\,\cite{wunsch,sarma}

\begin{equation}
\bar{\chi}_2(q,\,\omega)=\frac{e^2}{\epsilon_0q^2}\,\Pi_2(q,\,\omega)\ ,
\label{e20}
\end{equation}
where the non-retarded limit with $\omega\gg qc$ has been used and $\Pi_2(q,\,\omega)$ is defined as

\begin{equation}
\Pi_2(q,\,\omega)=\frac{4}{{\cal A}}\sum_{n_1,n_2,\,{\bf k}}\left.\left|\langle n_1,{\bf k}\right|e^{-i{\bf q}\cdot{\bf r}}\left.\right|n_2,{\bf k}+{\bf q}\rangle\right|^2\,
\frac{f_0(\varepsilon_{n_1,\,{\bf k}})-f_0(\varepsilon_{n_2,\,{\bf k}+{\bf q}})}{\varepsilon_{n_2,\,{\bf k}+{\bf q}}-\varepsilon_{n_1,\,{\bf k}}-\hbar(\omega+i0^+)}\ ,
\label{e21}
\end{equation}
${\cal A}$ is the surface area of a graphene micro-ribbon, $f_0(x)$
is the Fermi-Dirac function for thermal-equilibrium electrons,
$\varepsilon_{n,\,{\bf k}}$ is the kinetic energy of electrons, and
$n_1,\,n_2=\pm 1$ represent the valence ($-$) and conduction ($+$)
bands of Dirac cones in a graphene layer.
\medskip

The Dirac-cone model only holds for crystal momentum that is not too large.
In fact, the energy band structure ceases to be a cone above about 2.5 eV.
For our medium-high-density sample used in this paper, the Fermi momentum is much
less than the momentum separation between valleys at $K$ and $K^\prime$. The
Coulomb interaction between electrons in different valleys will be
significantly suppressed due to this large momentum separation. Additionally,
the electron-phonon mediated inter-valley scattering also requires a temperature
higher than   room temperature.  Since we assume very low temperature for electrons
in our model, this inter-valley scattering may be justifiably neglected.
\medskip

The 2D dielectric function $\kappa(q,\,\omega)$ for graphene micro-ribbons can be obtained by\,\cite{stern}
$\kappa(q,\,\omega)=\kappa_b+q\,\bar{\chi}_2(q,\,\omega)/2$ in the non-retarded limit with $\kappa_b$ being the dielectric constant of graphene materials.
In addition, the optical conductivity $\sigma_2(q,\,\omega)$ of graphene ribbons can be calculated from Eq.\,(\ref{e19}) within the linear-response theory,
i.e. $\sigma_2(q,\,\omega)=\omega\epsilon_0\,{\rm Im}[\bar{\chi}_2(q,\,\omega)]$.  In a similar way, $\sigma_1(q,\,\omega)$ can be calculated from Eq.\,(\ref{e128}) below.
Note that $q\,\bar{\chi}_2(q,\,\omega)$ along with Eq.\,(\ref{e20}) correspond to 2D rather than 3D Fourier transform of the Coulomb interaction.
\medskip

After a lengthy calculation\,\cite{oleksiy}, from Eq.\,(\ref{e21}) we get an analytical expression for a gapped graphene layer at $T=0$\,K as follows:

\begin{eqnarray}
\Pi_2(q,\,\omega)&=&\frac{2\mu_2}{\pi\hbar^2v_F^2}-\frac{q^2}{4\pi\hbar\sqrt{|v_F^2q^2-\omega^2|}}
\nonumber\\
&\times& \left\{i\left[G_>(x_{1,-})-G_>(x_{1,+})\right]{\cal Q}_{1_<}(x_{2,-})+\left[G_<(x_{1,-})+iG_>(x_{1,+})\right]{\cal Q}_{2_<}(x_{2,-},\,x_{2,+})\right.
\nonumber\\
&+& \left[G_<(x_{1,+})+G_<(x_{1,-})\right]{\cal Q}_{3_<}(x_{2,-})+\left[G_<(x_{1,-})-G_<(x_{1,+})\right]{\cal Q}_{4_<}(x_{2,+})
\nonumber\\
&+& \left[G_>(x_{1,+})-G_>(x_{1,-})\right]{\cal Q}_{1_>}(x_{2,-},\,x_3)+\left[G_>(x_{1,+})+iG_<(x_{1,-})\right]{\cal Q}_{2_>}(x_{2,-},\,x_{2,+})
\nonumber\\
&+& \left[G_>(x_{1,+})-G_>(-x_{1.-})-i\pi[2-x_0^2]\right]{\cal Q}_{3_>}(x_{2,+})
\nonumber\\
&+& \left[G_>(-x_{1,-})+G_>(x_{1,+})-i\pi[2-x_0^2]\right]{\cal Q}_{4_>}(x_{2,-},\,x_3)
\nonumber\\
&+&\left.\left[G_0(x_{1,+})-G_0(x_{1,-})\right]{\cal Q}_{5_>}(x_3)\right\}\ ,
\label{e22}
\end{eqnarray}
where $\mu_2=\sqrt{(\hbar v_Fk_{2F})^2+(E_G/2)^2}-E_G/2$ is the chemical potential of electrons
in a graphene layer with respect to zero energy at $k=0$, $v_F$ is the
Fermi velocity of graphene, the kinetic energy of electrons in
valence and conduction bands are

\begin{equation}
\varepsilon_{\pm,\,{\bf k}}=\pm\sqrt{\hbar^2v_F^2k^2+E_G^2/4}\ ,
\label{level}
\end{equation}
and $E_G$ is the induced bandgap. From Eq.\,(\ref{level}) we know that a finite effective mass $E_G/2v_F^2$ of electrons
will be created for a gapped graphene layer close to the edge of a gap with $k\ll E_G/\hbar v_F$.
\medskip

The coulomb interaction between electrons will induce a self-energy, leading to an energy-dependent Fermi velocity beyond the random-phase approximation\,\cite{arxiv}.
However, this energy-dependent behavior becomes significant only around the Dirac point\,\cite{elias}.
For our samples in this paper with a very high doping concentration, we expect this effect is minimized. Therefore, we have neglected the
coulomb renormalization to the graphene Fermi velocity in Eq.\,(\ref{level}).
Correction to the self-energy comes from  correlation effects. However,
this correction becomes significant only at low densities, as it occur,
for example, when there is  Wigner crystallization. However, the sample
considered in this paper is of medium-high-density concentration, where
correlation effects may be neglected.
Since we recognize that the limitations of the model is a poor argument when it comes to the experimental data,
we included in Eq.\,(24) an effective mass $\sim E_G$ term into Dirac fermion picture.
We note that the induced bandgap may
be attributed to either photon dressed-state effects by circularly polarized
light or to a substrate.
There exist two main sources of disorder in graphene ribbons, i.e., edge roughness and charged impurities.
Since wide micro-ribbons are considered in this paper, we can neglect the edge roughness. This leaves the charged impurities as the dominant scattering mechanism.
The existence of charged impurities is also able to
induce a self-energy, leading to a deformed density of states around the Dirac point of graphene\,\cite{hu}. However, for our medium-high-doping samples, the screening length
of impurity potential is small, leading to a negligent effect on the density of states in graphene.
Moreover, with the added gap in Eq.\,(\ref{level}), the charged impurities create localized states only in the range between  $0.92\,E_G$ and $E_G$\,\cite{gupta}.
Since our transmission is not substantially affected up to $E_G = 0.25 $\,eV (see Fig.\,11 in Sec.\,III)  and the induced states by localized impurities are well above the terahertz regime of interest in this paper,
we can neglect this effect here.
For impurities, the interference or localization effect on the conductivity
becomes important only for very high impurity concentrations.
Since the sample discussed in this paper is only a medium-high-density one,
localization is not expected to be a dominant effect in comparison with other
effects studied in this work.
The detailed discussion related to the screened disordered effect
on the electron polarizability of graphene beyond the random-phase approximation is, however, outside the scope of the current paper.
\medskip

Three functions introduced in Eq.\,(\ref{e22}) are defined as

\begin{eqnarray}
&&G_<(x)=x\sqrt{x_0^2-x^2}-\left(2-x_0^2\right)\,\cos^{-1}\left(\frac{x}{x_0}\right)\ , \label{e23}\\
&&G_>(x)=x\sqrt{x^2-x_0^2}-\left(2-x_0^2\right)\,\cosh^{-1}\left(\frac{x}{x_0}\right)\ , \label{e24}\\
&&G_0(x)=x\sqrt{x^2-x_0^2}-\left(2-x_0^2\right)\,\sinh^{-1}\left(\frac{x}{\sqrt{-x_0^2}}\right)\
. \label{e25}
\end{eqnarray}
Moreover, nine region functions used in Eq.\,(\ref{e22}) are defined by

\begin{eqnarray}
&&{\cal Q}_{1_<}(x_{2,-})=\theta(\mu_2-x_{2,-}-\hbar\omega)\ , \label{e26-1}\\
&&{\cal Q}_{2_<}(x_{2,-},\,x_{2,+})=\theta(-\hbar\omega-\mu_2+x_{2,-})\,\theta(\hbar\omega+\mu_2-x_{2,-})\,\theta(\mu_2+x_{2,+}-\hbar\omega)\ , \label{e26-2}\\
&&{\cal Q}_{3_<}(x_{2,-})=\theta(-\mu_2+x_{2,-}-\hbar\omega)\ , \label{e26-3}\\
&&{\cal Q}_{4_<}(x_{2,+})=\theta(\hbar\omega+\mu_2-x_{2,+})\,\theta(\hbar v_Fq-\hbar\omega)\ , \label{e26-4}\\
&&{\cal Q}_{1_>}(x_{2,-},\,x_3)=\theta(2k_{2F}-q)\,\theta(\hbar\omega-x_3)\,\theta(\mu_2+x_{2,-}-\hbar\omega)\ , \label{e26-5}\\
&&{\cal Q}_{2_>}(x_{2,-},\,x_{2,+})=\theta(\hbar\omega-\mu_2-x_{2,-})\,\theta(\mu_2+x_{2,+}-\hbar\omega)\ , \label{e26-6}\\
&&{\cal Q}_{3_>}(x_{2,+})=\theta(\hbar\omega-\mu_2-x_{2,+})\ , \label{e26-7}\\
&&{\cal Q}_{4_>}(x_{2,-},\,x_3)=\theta(q-2k_{2F})\,\theta(\hbar\omega-x_3)\,\theta(\mu_2+x_{2,-}-\hbar\omega)\ , \label{e26-8}\\
&&{\cal Q}_{5_>}(x_3)=\theta(\hbar\omega-\hbar v_Fq)\,\theta(x_3-\hbar\omega)\ , \label{e26-9}
\end{eqnarray}
where $k_{2F}=\sqrt{(\mu_2+E_G/2)^2-(E_G/2)^2}/\hbar v_F$ is the Fermi wave number and
$\theta(x)$ is the unit step function. Finally, we have defined six
variables $x_0,\,x_{1,\pm},\,x_{2,\pm}$ and $x_3$ through

\begin{eqnarray}
&&x_0=\sqrt{1+\frac{E_G^2}{\hbar^2v_F^2q^2-\hbar^2\omega^2}}\ , \label{e27}\\
&&x_{1,\pm}=\frac{2\mu_2\pm\hbar\omega}{\hbar v_Fq}\ , \label{e28}\\
&&x_{2,\pm}=\sqrt{\hbar^2v_F^2(q\pm k_{2F})^2+E_G^2/4}\ , \label{e29}\\
&&x_3=\sqrt{\hbar^2v_F^2q^2+E_G^2}\ . \label{e30}
\end{eqnarray}
\medskip

Following a similar approach, within the random-phase approximation we get
the nonlocal optical response function for the 2DEG inside an InAs QW\,\cite{stern}

\begin{eqnarray}
\bar{\chi}_1(q,\,\omega)&=&\frac{2\rho_se^2m_s^{\ast}}{\epsilon_0\hbar^2k_{1F}q^3}
\left\{\left[2z-C_-\sqrt{(z-u)^2-1}-C_+\sqrt{(z+u)^2-1}\,\right]\right.
\nonumber\\
&+&\left.i\left[D_-\sqrt{1-(z-u)^2}-D_+\sqrt{1-(z+u)^2}\,\right]\right\}
\ , \label{e128}
\end{eqnarray}
where $\rho_s$ is the 2DEG density, $k_{1F}=\sqrt{2\pi\rho_s}$ and $m_s^{\ast}$ are
the Fermi wave number and the
effective mass of 2DEG. Additionally, we have defined the notations in
Eq.\,(\ref{e128}) for a driven 2DEG: $u=m_s^{\ast}\omega/\hbar
qk_{1F}$, $z=q/2k_{1F}$, $C_+=(z+u)/|z+u|$ and $D_+=0$ ($C_+=0$ and
$D_+=1$) for $|z+u|>1$ ($|z+u|<1$), and $C_-=(z-u)/|z-u|$ and
$D_-=0$ ($C_-=0$ and $D_-=1$) for $|z-u|>1$ ($|z-u|<1$).
\medskip

For the specular mode with $n=0$, from Eqs.\,(\ref{e11})--(\ref{e14}) the boundary conditions for the
EM fields at the interface yield

\begin{equation}
\left.B_x(q_x\right|\omega)-\left.C_x(q_x\right|\omega)
=-\left.A_x(q_x\right|\omega)\ ,
\label{e31}
\end{equation}

\begin{equation}
\left.B_y(q_x\right|\omega)-\left.C_y(q_x\right|\omega)=-\left.A_y(q_x\right|\omega)\ ,
\label{e32}
\end{equation}

\[
-\frac{ic^2\beta_{1,0}}{\omega^2}\left.B_y(q_x\right|\omega)-\frac{ic^2\beta_{2,0}}{\omega^2}
\left.C_y(q_x\right|\omega)-\zeta\sum\limits_{n^\prime=-\infty}^{\infty}{\rm
sinc}(n^{\prime}\pi\zeta)\,\bar{\chi}_2(q_{n^\prime},\,\omega)\left.B_y(q_{n^\prime}\right|\omega)
\]
\begin{equation}
-\bar{\chi}_1(q_x,\,\omega)\left.B_y(q_x\right|\omega)=\left(\bar{\chi}_1(q_x,\,\omega)+\zeta\,\bar{\chi}_2(q_x,\,\omega)-\frac{ic^2\eta}{\omega^2}\right)\left.A_y(q_x\right|\omega)\
, \label{e33}
\end{equation}

\[
-\frac{i}{\beta_{1,0}}\,n^2_a\left.B_x(q_x\right|\omega)-\frac{i}{\beta_{2,0}}\,\epsilon_s(q_x,\beta_{2,0},\,\omega)\left.C_x(q_x\right|\omega)
\]
\[
-\zeta\sum\limits_{n^\prime=-\infty}^{\infty}{\rm
sinc}(n^{\prime}\pi\zeta)\,\bar{\chi}_2(q_{n^\prime},\,\omega)\left.B_x(q_{n^\prime}\right|\omega)-\bar{\chi}_1(q_x,\,\omega)\left.B_x(q_x\right|\omega)
\]
\begin{equation}
=\left(\bar{\chi}_1(q_x,\,\omega)+\zeta\,\bar{\chi}_2(q_x,\,\omega)-
\frac{i}{\eta}\right)\left.A_x(q_x\right|\omega)\ .
\label{e34}
\end{equation}
The terms with $n^\prime\neq 0$ in Eqs.\,(\ref{e33}) and (\ref{e34})
reflect the coupling between the specular mode and higher-order
diffraction modes. These generated higher-order diffracted EM fields can
be treated as source terms in addition to the incident EM field in a
perturbation picture. If we exclude all the terms with $n^\prime\neq
0$ in Eqs.\,(\ref{e33}) and (\ref{e34}), the solution of
Eqs.\,(\ref{e31})--(\ref{e34}) gives rise to a transverse
EM response for a single graphene sheet by taking
$\bar{\chi}_1(q_x,\,\omega)=0$, $\zeta=1$ and
$\epsilon_s(q_x,\beta_{2,0},\,\omega)=\epsilon_b$\,\cite{ziegler,soljacic}.
\medskip

For the diffraction modes with $n=\pm 1,\,\pm 2,\,\cdots$, on the other hand, from Eqs.\,(\ref{e11})--(\ref{e14}) the boundary
conditions for the EM fields lead to

\begin{equation}
\left.B_x( q_n\right|\omega)-\left.C_x( q_n\right|\omega)=0\ ,
\label{e35}
\end{equation}

\begin{equation}
\left.B_y(q_n\right|\omega)-\left.C_y(q_n\right|\omega)=0\ , \label{e36}
\end{equation}

\[
-\frac{ic^2\beta_{1,n}}{\omega^2}\left.B_y(q_n\right|\omega)
-\frac{ic^2\beta_{2,n}}{\omega^2}\left.C_y(q_n\right|\omega)
\]
\begin{equation}
-\zeta\sum_{n^\prime =-\infty}^{\infty}{\rm
sinc}[(n^{\prime}-n)\pi\zeta]\,\bar{\chi}_2(q_{n^\prime},\,\omega)\left.B_y(q_{n^\prime}\right|\omega)
-\bar{\chi}_1(q_n,\,\omega)\left.B_y(q_n\right|\omega)=0\ ,
\label{e37}
\end{equation}

\[
-\frac{i}{\beta_{1,n}}\,n_a^2\left.B_x(q_n\right|\omega)-\frac{i}{\beta_{2,n}}\,\epsilon_s(q_n,\beta_{2,n},\,\omega)\left.C_x(q_n\right|\omega)
\]
\begin{equation}
-\zeta\sum_{n^\prime =-\infty}^{\infty}{\rm
sinc}[(n^{\prime}-n)\pi\zeta]\,\bar{\chi}_2(q_{n^\prime},\,\omega)\left.B_x(q_{n^\prime}\right|\omega)
-\bar{\chi}_1(q_n,\,\omega)\left.B_x(q_n\right|\omega)=0\ , \label{e38}
\end{equation}
where $\bar{\chi}_1=0$ and $\bar{\chi}_1\neq 0$ correspond to a GMRA
only or a 2DEG plus a GMRA, respectively. The terms with
$n^\prime=0$ in Eqs.\,(\ref{e37}) and (\ref{e38}) reflect the
coupling between different diffraction modes with the specular mode.
Here, the terms with $n^\prime=0$, excited directly by the incident
EM field, can be regarded as source terms for the generated higher-order
diffracted EM fields with $n=\pm 1,\,\pm 2,\,\cdots$ in a perturbation
picture.
\medskip

Once the solution of the matrix equation derived from Eqs.\,(\ref{e31})--(\ref{e38}) (see Appendix\ A) is obtained, for $p$ polarization
we can express the intensities of EM field at the interface as

\begin{eqnarray}
\left.|E_x(x,z=0\right|\omega)|^2&=&\left.[A_x(q_x\right|\omega)]^2+\sum\limits_{n=-\infty}^{\infty}\left.|B_x(q_n\right|\omega)|^2
+\left.2A_x(q_x\right|\omega)\,{\rm Re}\left.\left[B_x(q_x\right|\omega)\right]
\nonumber\\
&+&\sum\limits_{n,\,n^{\prime}}^{(n\neq
n^{\prime})}\left.B_x(q_n\right|\omega)
\left.[B_x(q_{n^{\prime}}\right|\omega)]^{\ast}\,e^{i(n-n^{\prime})Gx}
\nonumber\\
&+&\left.2A_x(q_x\right|\omega)\,{\rm Re}\,\left(\sum_{n\neq
0}\left.B_x(q_n\right|\omega)\,e^{inGx}\right)\ , \label{e39}
\end{eqnarray}

\begin{eqnarray}
\left.|E_y(x,z=0\right|\omega)|^2&=&\left.[A_y(q_x\right|\omega)]^2+\sum\limits_{n=-\infty}^{\infty}\left.|B_y(q_n\right|\omega)|^2
+\left.2A_y(q_x\right|\omega)\,{\rm Re}\left.\left[B_y(q_x\right|\omega)\right]
\nonumber\\
&+&\sum\limits_{n,\,n^{\prime}}^{(n\neq
n^{\prime})}\left.B_y(q_n\right|\omega)\left.[B_y(q_{n^{\prime}}\right|\omega)]^{\ast}\,e^{i(n-n^{\prime})Gx}
\nonumber\\
&+&\left.2A_y(q_x\right|\omega)\,{\rm Re}\,\left(\sum\limits_{n\neq
0}\left.B_y(q_n\right|\omega)\,e^{inGx}\right)\ , \label{e40}
\end{eqnarray}

\[
\left.|E_z(x,z=0\right|\omega)|^2=\frac{q_x^2}{\eta^2}
\left.[A_x(q_x\right|\omega)]^2+\sum\limits_{n=-\infty}^{\infty}\,
\frac{q_n^2}{|\beta_{1,n}|^2}\left.|B_x(q_n\right|\omega)|^2
\]
\[
-\frac{2q_x^2}{\eta^2}\left.A_x(q_x\right|\omega)\,{\rm
Re}\left.\left[B_x(q_x\right|\omega)\right]
+\sum\limits_{n,\,n^{\prime}}^{(n\neq
n^{\prime})}\,\frac{q_nq_{n^{\prime}}}
{\beta_{1,n}(\beta_{1,n^{\prime}})^{\ast}}
\left.B_x(q_n\right|\omega)\left.[B_x(q_{n^{\prime}}\right|\omega)]^{\ast}
\]
\begin{equation}
\times e^{i(n-n^{\prime})Gx}-\left.2A_x(q_x\right|\omega)\,{\rm
Re}\,\left(\sum\limits_n^{(n\neq 0)}\,\frac{q_xq_n}{\eta\beta_{1,n}}
\left.B_x(q_n\right|\omega)\,e^{inGx}\right)\ , \label{e41}
\end{equation}
where the interferences taking place in the $x$ direction at the interface between the reflected EM field ($n\neq n^\prime$)
and that between the incident and reflected EM fields ($n\neq 0$) are all included.
\medskip

Using Eqs.\,(\ref{e39})--(\ref{e41}),
we get the $x$-dependent square of the ratio $\left.{\cal R}(x\right|\omega)$ of the reflected to the
incident ${\bf E}$-field amplitude at the interface, given by

\begin{equation}
\left.{\cal R}^{\gtrless}(x\right|\omega)=\frac{\left.|E^r_x(x,z=0\right|\omega)|^2+\left.|E^r_y(x,z=0\right|\omega)|^2+\left.|E^r_z(x,z=0\right|\omega)|^2}
{(1+q_x^2/\eta^2)\left.A^2_x(q_x\right|\omega)+\left.A^2_y(q_x\right|\omega)}\ ,
\label{e42}
\end{equation}
where $\left.{\cal R}^{\gtrless}(x\right|\omega)$ contains both the near ($<$) and far ($>$) field contributions, and

\begin{equation}
\left.|E^r_x(x,z=0\right|\omega)|^2=\sum\limits_{n,\,n^{\prime}}\left.B_x(q_n\right|\omega)
\left.[B_x(q_{n^{\prime}}\right|\omega)]^{\ast}\,e^{i(n-n^{\prime})Gx}\ , \label{e43}
\end{equation}

\begin{equation}
\left.|E^r_y(x,z=0\right|\omega)|^2=\sum\limits_{,\,n^{\prime}}\left.B_y(q_n\right|\omega)\left.[B_y(q_{n^{\prime}}\right|\omega)]^{\ast}\,e^{i(n-n^{\prime})Gx}\ ,
\label{e44}
\end{equation}

\begin{equation}
\left.|E^r_z(x,z=0\right|\omega)|^2=\sum\limits_{n,\,n^{\prime}}\,\frac{q_nq_{n^{\prime}}}
{\beta_{1,n}(\beta_{1,n^{\prime}})^{\ast}}\,
B_x(q_n;\,\omega)\left.[B_x(q_{n^{\prime}}\right|\omega))]^{\ast}\,e^{i(n-n^{\prime})Gx}\
. \label{e45}
\end{equation}
Following the same approach, we get the $x$-dependent square of
the ratio $\left.{\cal F}(x\right|\omega)$ of the transmitted to the
incident ${\bf E}$-field amplitude at the interface, given by

\begin{equation}
\left.{\cal F}^{\gtrless}(x\right|\omega)=\frac{\left.|E^t_x(x,z=0\right|\omega)|^2+\left.|E^t_y(x,z=0\right|\omega)|^2+\left.|E^t_z(x,z=0\right|\omega)|^2}
{(1+q_x^2/\eta^2)\left.A^2_x(q_x\right|\omega)+\left.A^2_y(q_x\right|\omega)}\ ,
\label{e46}
\end{equation}
where $\left.{\cal F}^{\gtrless}(x\right|\omega)$ also includes both the near ($<$) and far ($>$) field contributions, and

\begin{equation}
\left.|E^t_x(x,z=0\right|\omega)|^2=\sum\limits_{n,\,n^{\prime}}\left.C_x^T(q_n\right|\omega)
\left.[C_x(q_{n^{\prime}}\right|\omega)]^{\ast}\,e^{i(n-n^{\prime})Gx}\ , \label{e47}
\end{equation}

\begin{equation}
\left.|E^t_y(x,z=0\right|\omega)|^2=\sum\limits_{n,\,n^{\prime}}\left.C_y(q_n\right|\omega)\left.[C_y(q_{n^{\prime}}\right|\omega)]^{\ast}\,e^{i(n-n^{\prime})Gx}\ ,
\label{e48}
\end{equation}

\begin{equation}
\left.|E^t_z(x,z=0\right|\omega)|^2=\sum\limits_{n,\,n^{\prime}}\,
\frac{q_nq_{n^\prime}}{\beta_{2,n}(\beta_{2,n^\prime})^\ast}\left.C_x(q_n\right|\omega)
\left.[C_x(q_{n^{\prime}}\right|\omega)]^{\ast}\,e^{i(n-n^{\prime})Gx}\
. \label{e49}
\end{equation}
After integrating Eqs.\,(\ref{e42}) and (\ref{e46}) with respect to
$x$, we obtain the far-field reflectivity and transmissivity
spectra of the sample as

\begin{equation}
{\cal R}(\omega)=\lim\limits_{L_x\to\infty}\,\frac{1}{2L_x}\,\int\limits_{-L_x}^{L_x} dx\left.{\cal R}^>(x\right|\omega)
=\sum\limits_{n=-\infty}^{\infty}\left.R_n(q_n\right|\omega)\,\theta(n_a\,\omega-|q_n|c)\ ,
\label{e50}
\end{equation}

\begin{equation}
{\cal F}(\omega)=\lim\limits_{L_x\to\infty}\,\frac{1}{2L_x}\,\int\limits_{-L_x}^{L_x} dx\left.{\cal F}^>(x\right|\omega)
=\sum\limits_{n=-\infty}^{\infty}\left.F_n(q_n\right|\omega)\,\theta(\sqrt{\epsilon^\prime_s}\,\omega-|q_n|c)\,\theta(\epsilon^\prime_s)\ ,
\label{e51}
\end{equation}
where $\epsilon^\prime_s\equiv{\rm Re}(\epsilon_s)$, the interference terms between different diffraction modes are canceled out and

\begin{equation}
\left.R_n(q_n\right|\omega)=\frac{(1+q_n^2/|\beta_{1,n}|^2)\left.|B_x(q_n\right|\omega)|^2+\left.|B_y(q_n\right|\omega)|^2}
{(1+q_x^2/\eta^2)\left.A^2_x(q_x\right|\omega)+\left.A^2_y(q_x\right|\omega)}\
, \label{e52}
\end{equation}

\begin{equation}
\left.F_n(q_n\right|\omega)=\frac{(1+|q_n^2/|\beta_{2,n}|^2)\left.|C_x(q_n\right|\omega)|^2+\left.|C_y(q_n\right|\omega)|^2}
{(1+q_x^2/\eta^2)\left.A^2_x(q_x\right|\omega)+\left.A^2_y(q_x\right|\omega)}\
. \label{e53}
\end{equation}

\section{Numerical Results and Discussions}
\label{sect3}

In our numerical calculations, we assume $p$ polarization
for the incident EM-field and set the parameters in numerical
calculations as follows: $\rho_s=8\times 10^{11}$\,cm$^{-2}$,
$m_s^\ast=0.024\,m_0$ with $m_0$ being the free electron mass,
$E_G=0$, $\mu_2=0.45$\,eV, $v_F=10^6$\,m/sec,
$\epsilon_b=12$, $\hbar\Omega_{\rm pl}=10.14$\,meV,
$\hbar\gamma_0=5$\,meV, $\zeta=0.5$, $d=4\,\mu$m (deep sub-wavelength regime),
$\theta_i=30^{\rm o}$ and $q_x=(\omega/c)\,n_a\sin\theta_i$ with $n_a=1$. In this paper, we choose the color scale from blue (darker, minimum) to red (lighter, maximum).
The change of these parameters will be
directly indicated in the figure captions. In this paper, we only consider the low-energy intraband plasmon excitation in the terahertz-frequency range.
\medskip

\begin{figure}[htbp]
%\centerline{\epsfig{file=figure2.eps,width=3.0in}}
\centerline{\includegraphics[width=3.0in]{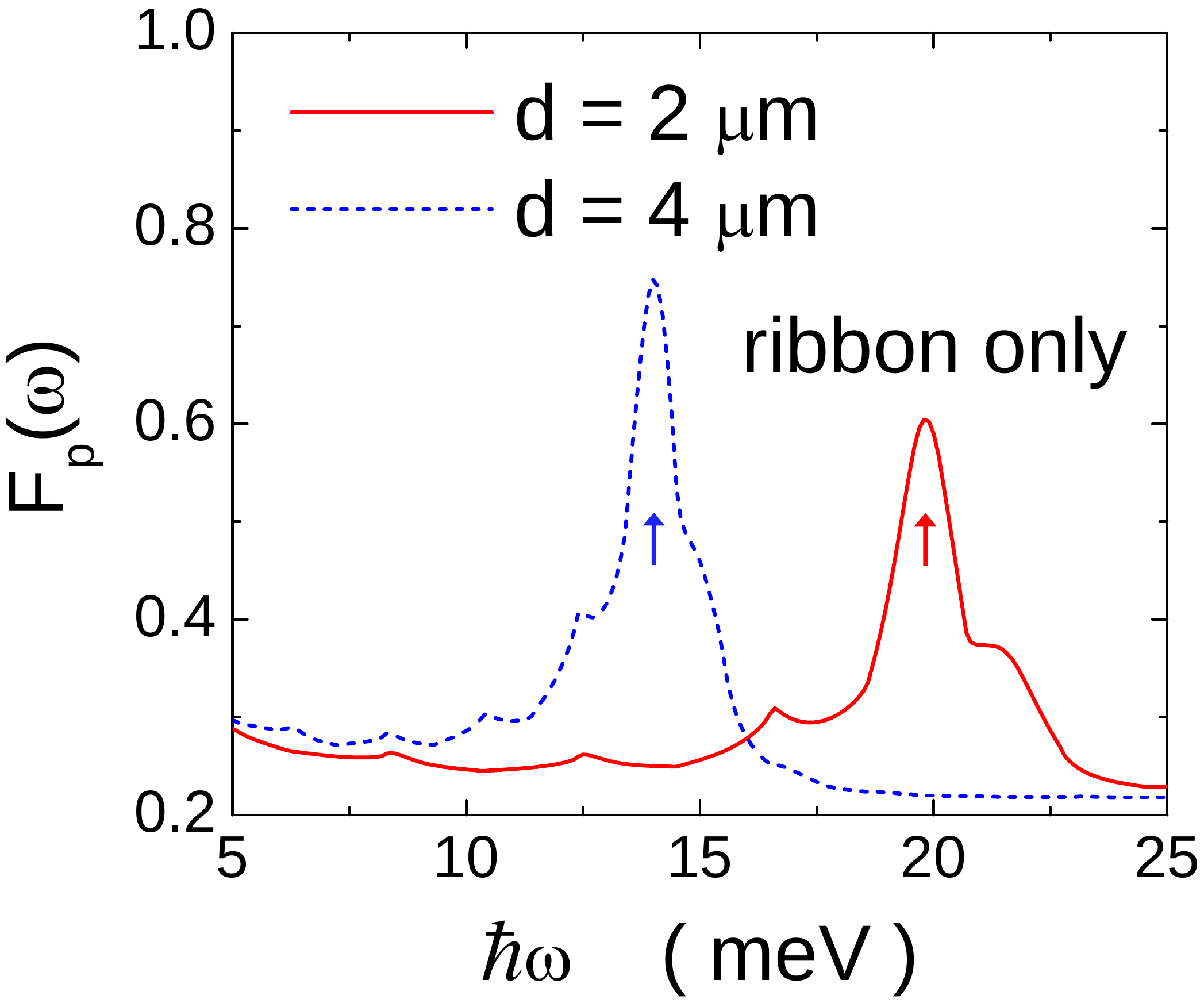}}
\caption{(Color online) A comparison of the far-field transmissivity spectra
${\cal F}_p(\omega)$ in the absence of both SP and QW for $p$ polarization with two given linear-array
periods: $d=2\,\mu$m (red solid curve) and $d=4\,\mu$m (blue dashed curve).
Two arrows indicate the shift of two corresponding peaks with $d$.
Other parameters in calculations are given in the text.} \label{f2}
\end{figure}

To uncover the physical mechanism behind the reported
graphene-plasmon peak shift with GMRA period\,\cite{ju}, we show a
comparison in Fig.\,\ref{f2} for two calculated far-field
transmissivity spectra ${\cal F}_p(\omega)$, defined by
Eq.\,(\ref{e51}), as a function of photon energy $\hbar\omega$ with
$d=2\,\mu$m (red solid curve) and $d=4\,\mu$m (blue dashed curve).
Here, only the graphene micro-ribbon array is included in our
system. As $d$ increases, the graphene plasmon peak at
$\hbar\omega=19.8$\,meV for $d=2\,\mu$m shifts down to
$\hbar\omega=14.0$\,meV for $d=4\,\mu$m and this peak shift
accurately satisfies the $\sim 1/\sqrt{d}$ scaling relation
experimentally observed\,\cite{ju}. A series of kinks below the
plasmon peak in Fig.\,\ref{f2} corresponds to the single-particle
excitations at $\hbar\omega=\hbar v_F|q_n|$ (for $|q_n|/k_{2F}<1$)
with $q_n\approx n(2\pi/d)$ for different values of integer $n$. On
the other hand, the shoulder appearing above the dominant peak comes
from another severely-Landau-damped graphene-plasmon peak, which
becomes more visible for $d=2\,\mu$m since the peak separation is
increased.
\medskip

\begin{figure}[htbp]
%\centerline{\epsfig{file=figure3.eps,width=3.0in}}
\centerline{\includegraphics[width=3.0in]{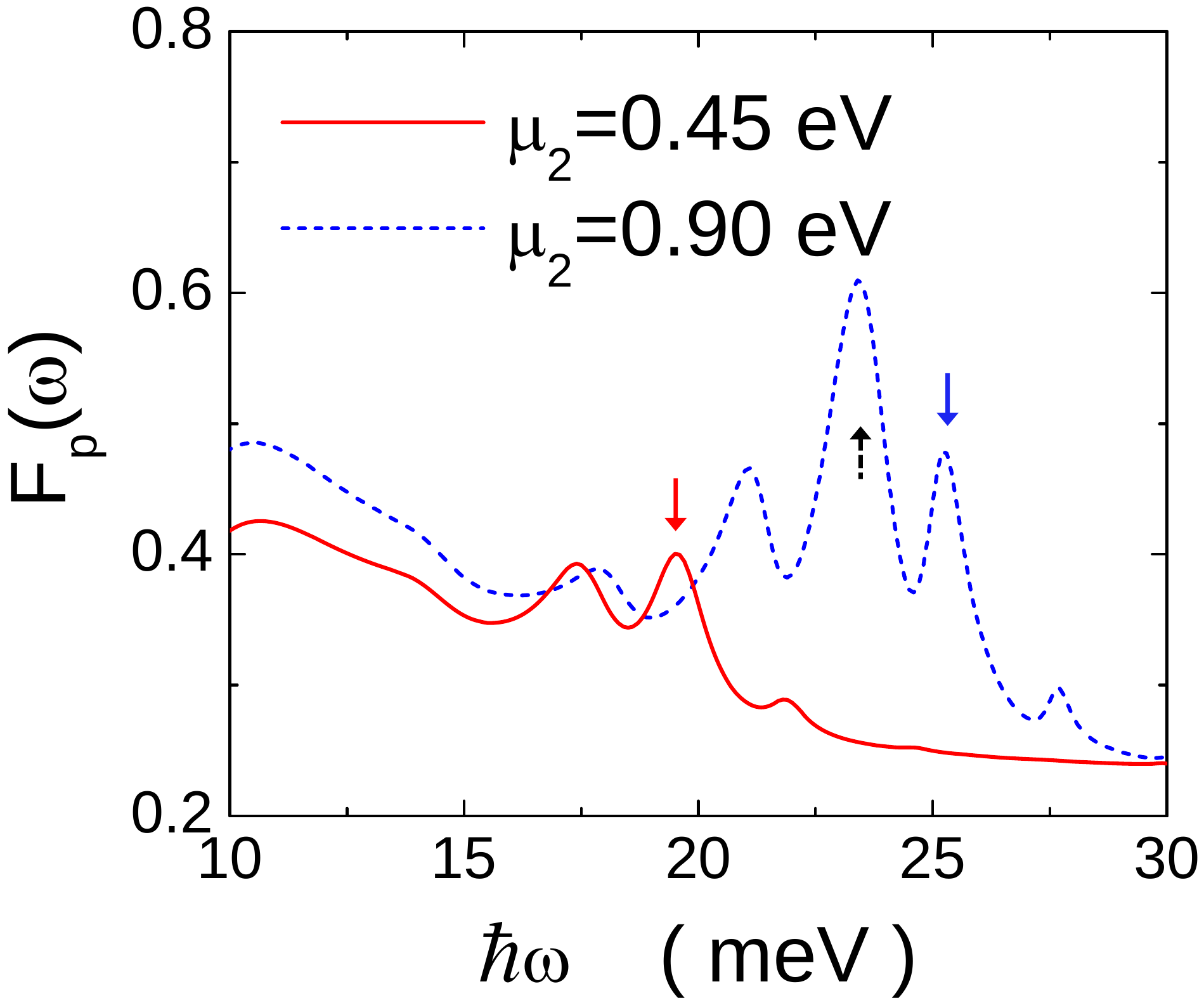}}
\caption{(Color online) A comparison of the calculated transmissivity
spectra ${\cal F}_p(\omega)$ for $p$ polarization with
$\mu_2=0.45$\,eV (red solid curve) and $\mu_2=0.9$\,eV (blue dashed curve).
Two downward solid-line arrows indicate the shift of two corresponding peaks with $\mu_2$,
while one upward dashed-line arrow indicates a new peak.
Other parameters in calculations are given in the text.} \label{f3}
\end{figure}

To see the many-body effect on the hybridized GMRA-QW plasmon excitation,
Fig.\,\ref{f3} presents a comparison for ${\cal F}_p(\omega)$ as a function of $\hbar\omega$
with two values for the graphene chemical potential,
i.e., $\mu_2=0.45$\,eV (red solid curve) and $\mu_2=0.9$\,eV (blue dashed curve).
Here, the chemical potential $\mu_2$ of graphene ribbons can be tuned by a gate voltage\,\cite{ju}.
As expected, one of the two graphene-like plasmon peaks at $\hbar\omega=19.5$\,meV for $\mu_2=0.45$\,eV moves up to $\hbar\omega=25.2$\,meV for $\mu_2=0.9$\,eV.
Interestingly, from this hybridized GMRA-QW plasmon peak shift we find that the plasmon energy is not scaled as $\sim\sqrt{\mu_2}$,
which is different from the observation reported in Ref.\,\onlinecite{ju} for the ribbon-only case. However, our calculated graphene plasmon peak shift
(from $14.0$\,meV to $19.9$\,meV) for the ribbon-only case
(not shown here) fully agrees with their observation\,\cite{ju}.
The $\mu_2$-insensitive plasmon peak at $\hbar\omega=10.7$\,meV is associated with the SP excitation at the photon energy
$\hbar\omega=\sqrt{\epsilon_b/(1+\epsilon_b)}\,\hbar\Omega_{\rm pl}\approx 10$\,meV
and will be addressed in Fig.\,\ref{f6} below.
Additionally, when $\mu_2=0.9$\,eV, we see a new strong graphene-like plasmon peak show up at $\hbar\omega=23.4$\,meV
because one relatively-flattened portion\,\cite{wunsch,oleksiy} on the plasmon dispersion curve at a relative large $|q_n|$ value becomes free of Landau-damping.
\medskip

\begin{figure}[htbp]
%\centerline{\epsfig{file=figure4a.eps,width=3.0in}}
%\epsfig{file=figure4b.eps,width=3.0in}
\centerline{\includegraphics[width=3.0in]{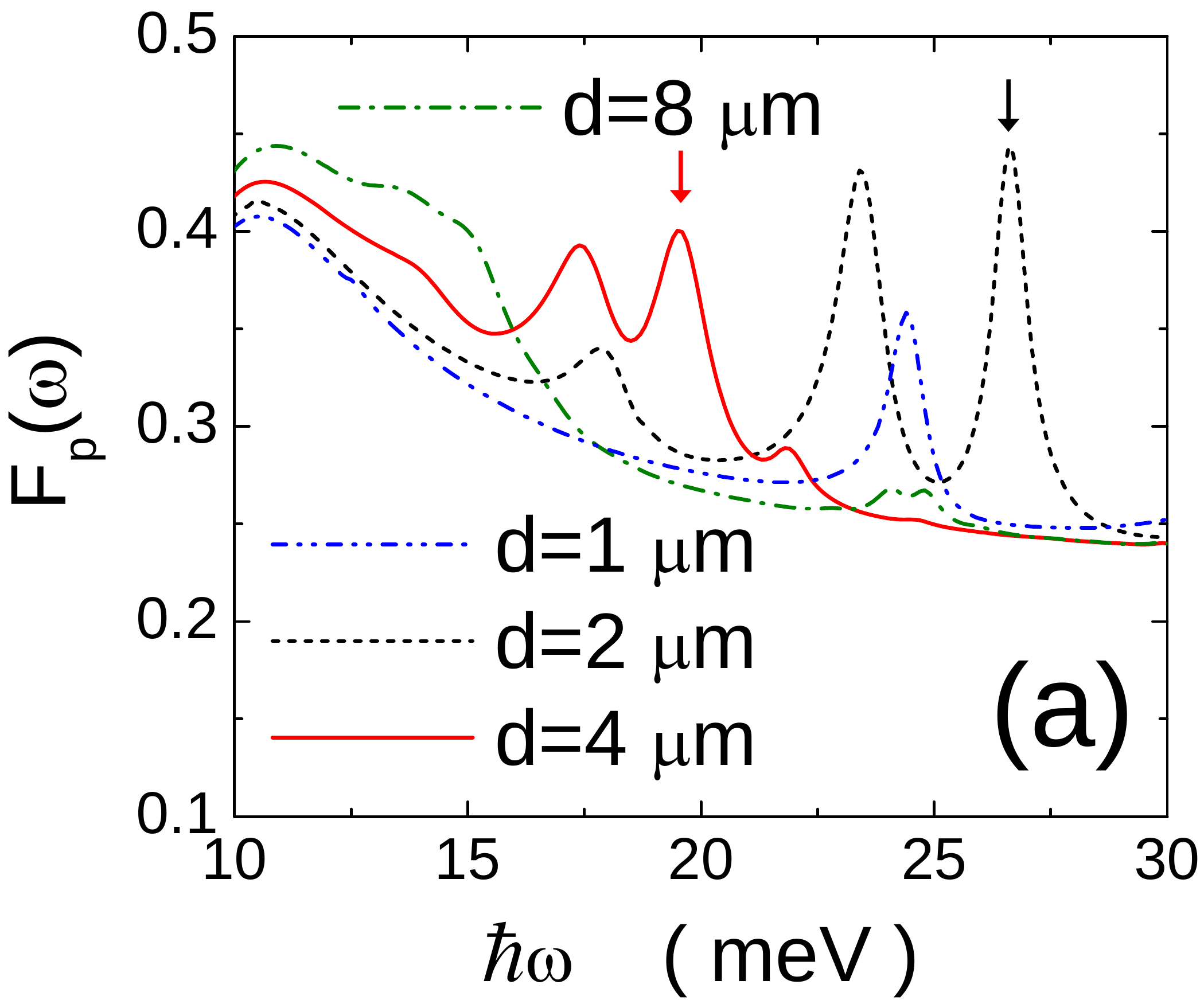}}
\centerline{\includegraphics[width=3.0in]{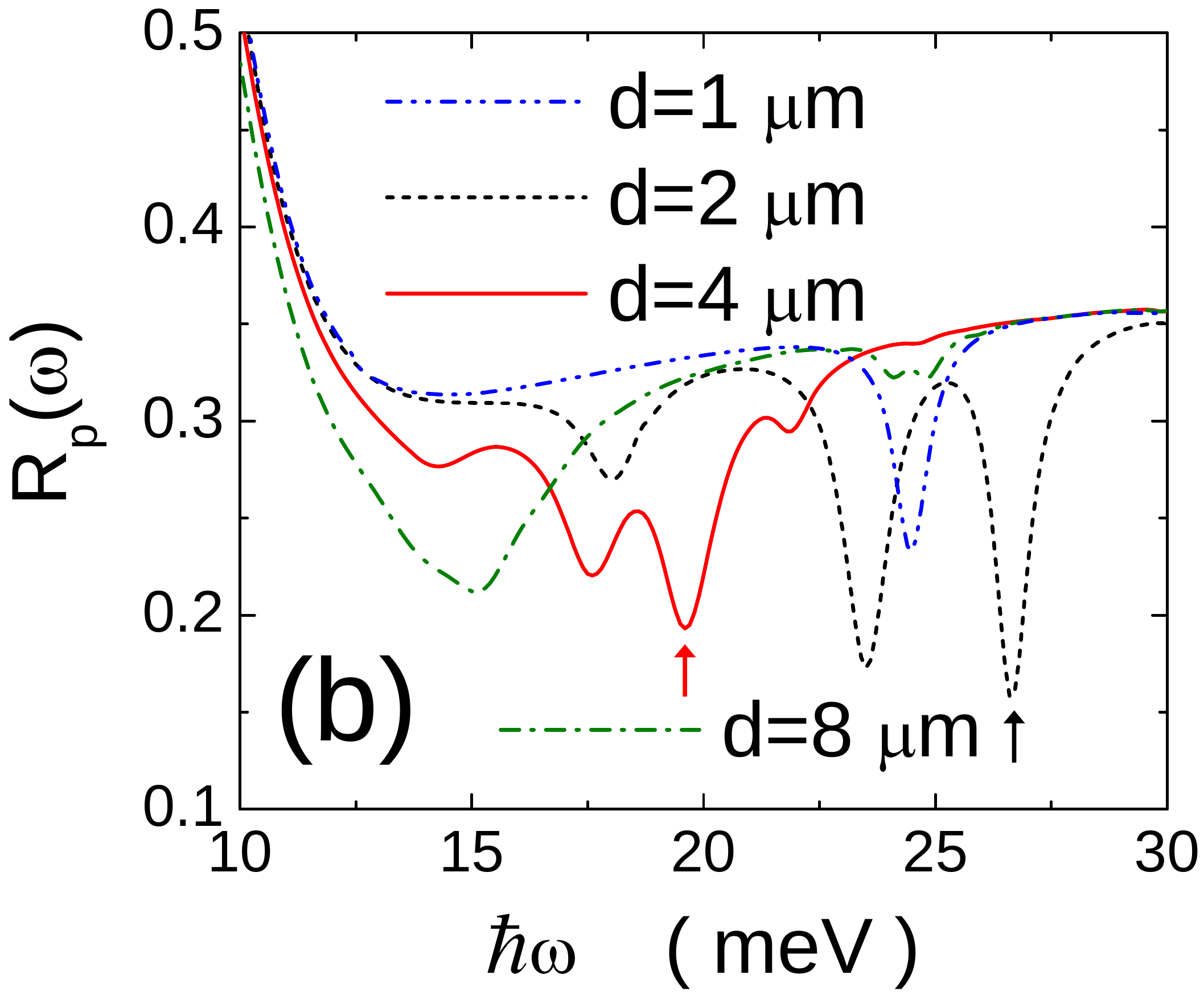}}
\caption{(Color online) Comparisons of the far-field transmissivity spectra ${\cal
F}_p(\omega)$ [in (a)] and the far-field reflectivity spectra ${\cal R}_p(\omega)$ [in
(b)] for $p$ polarization with different linear-array
periods: $d=1\,\mu$m (blue dash-dot-dotted curves), $d=2\,\mu$m (black dashed curves),
$d=4\,\mu$m (red solid curves) and $d=8\,\mu$m (green dash-dotted curves).
Two arrows indicate the shift of two corresponding peaks with $d$.
Other parameters in calculations are given in the text.} \label{f4}
\end{figure}

To further elucidate the importance of the EM coupling between the graphene plasmon in GMRA, as shown in Fig.\,\ref{f2}, and the 2DEG plasmon in a QW,
we displayed in Fig.\,\ref{f4} comparisons
for ${\cal F}_p(\omega)$ [in (a)] and for the far-field reflectivity spectra ${\cal R}_p(\omega)$, defined by Eq.\,(\ref{e50}), [in (b)] as functions of $\hbar\omega$
by choosing four different array periods: $d=1\,\mu$m (blue dash-dot-dotted curves), $d=2\,\mu$m (black dashed curves), $d=4\,\mu$m (red solid curves) and $d=8\,\mu$m (green dash-dotted curves).
Here, we fixed the ratio of the ribbon width ${\cal W}$ to the array period $d$ by $\zeta={\cal W}/d=0.5$.
As indicated by two downward arrows in Fig.\,\ref{f4}(a) for the shift of one of the two graphene-like plasmon peaks from $\hbar\omega=19.5$\,meV for $d=4\,\mu$m to $\hbar\omega=26.6$\,meV for $d=2\,\mu$m,
we determine that the hybridized GMRA-QW plasmon energy is approximately proportional to $1/\sqrt{d}\sim 1/\sqrt{{\cal W}}$ for $\zeta=0.5$.
When $d=8\,\mu$m, only two weakly-split plasmon peaks are visible at $\hbar\omega=24.1$\,meV and $\hbar\omega=24.8$\,meV, respectively.
As $d$ decreases from $2\,\mu$m to $1\,\mu$m,
the peak at $\hbar\omega=17.8$\,meV for $d=2\,\mu$m is pushed up to $\hbar\omega=24.4$\,meV.
These peak shifts with GMRA period are also reflected in
Fig.\,\ref{f4}(b), as indicated by the two upward arrows for example, and the steep rise of ${\cal R}_p(\omega)$
around $\hbar\omega=10$\,meV is associated with the excitation of SP mode in the system.
\medskip

\begin{figure}[htbp]
%\centerline{\epsfig{file=figure5.eps,width=3.0in}}
\centerline{\includegraphics[width=3.0in]{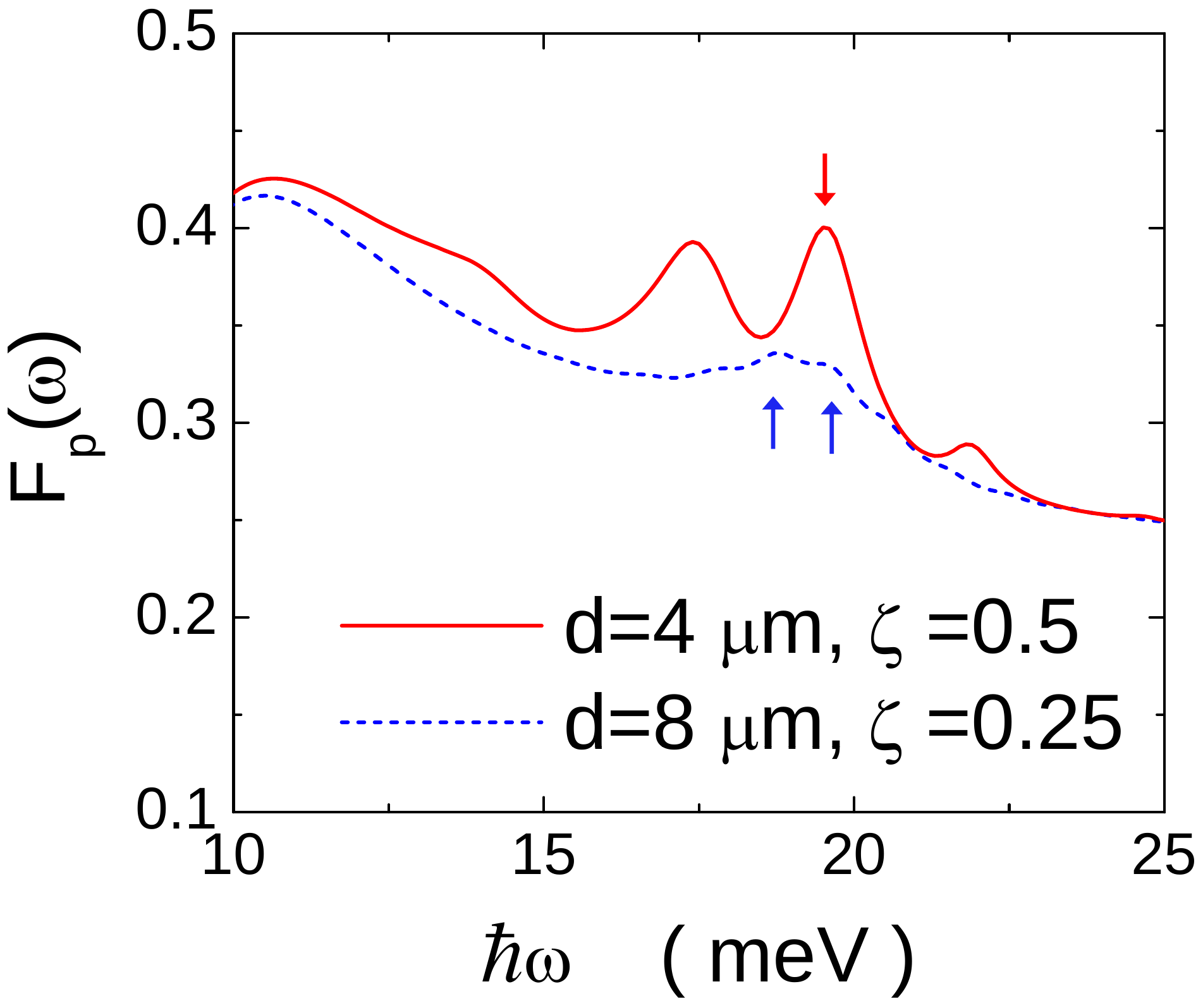}}
\caption{(Color online) A comparison of the calculated
transmissivity spectra ${\cal F}_p(\omega)$ for $p$ polarization
with $d=4\,\mu$m and $\zeta=0.5$ (red solid curve) as well as with
$d=8\,\mu$m and $\zeta=0.25$ (blue dashed curve). The peak indicated
by a downward red arrow splits into two indicated by two upward blue
arrows. Other parameters in calculations are given in the text.}
\label{f5}
\end{figure}

In order to clarify the fact that the graphene-like plasmon energy is
actually proportional to $1/\sqrt{{\cal W}}$, due to the
non-locality in the optical response of interacting electrons, instead of
$1/\sqrt{d}$ in single-particle modes, we show ${\cal F}_p(\omega)$ in
Fig.\,\ref{f5} for two cases with the same value for ${\cal W}$ but
different values for $d$. For the case with $\zeta=0.25$ and
$d=8\,\mu$m, we find a weakly-split peak (indicated by two upward arrows), which is aligned with a
strong non-split peak (indicated by a downward arrow) at $\zeta=0.5$ and $d=4\,\mu$m. This clearly
demonstrates that the graphene-like plasmon peak energy is proportional to
$1/\sqrt{{\cal W}}$ instead of $1/\sqrt{d}$, which is again in
agreement with the observation by Ju {\em et al\/}\,\cite{ju}.
The finite width of a ribbon introduces a characteristic wavenumber
(proportional to $1/{\cal W}$), which enforces a cut-off to the
short-range coulomb interaction between electrons in a micro-ribbon. A
smaller $\zeta$ value implies a weaker coupling between the 2DEG
and graphene plasmon excitations. The splitting of the
plasmon peak for $\zeta=0.25$ is attributed to the
contribution of even-integer diffraction modes,
while only odd-integer diffraction modes can contribute for
$\zeta=0.5$, as can be verified by the ${\rm sinc}(n^\prime\pi\zeta)$--terms appearing in Eqs.\,(\ref{e33}) and (\ref{e34}).
\medskip

Figure\ \ref{f6} presents in the left panel the individual and
combined effects of SP, 2DEG and GMRA as well as the strong EM
coupling among these plasmon excitations. The periodicity of GMRA
opens a minigap between any two adjacent branches (Bloch modes) in a
folded hybridized-plasmon dispersion curve. These minigaps can be
opened either at the Brillouin-zone center or at the Brillouin-zone
boundary. In the deep sub-wavelength regime, only the minigap at the
Brillouin-zone center leads to a peak in the far-field
transmissivity spectrum\,\cite{aam1}. When the doped GaAs substrate
is changed into an undoped one (no SP, blue dash-dot-dotted curve),
only 2DEG and graphene plasmons can exist in this system. In this
case, there exists strong EM coupling between the graphene and 2DEG
plasmon excitations. As labeled by the circled numbers, peak--1 and
peak--4 are related to the 2DEG-like plasmons, while peak--2 and
peak--3 are connected to the graphene-like plasmon excitations. Two
small kinks on the outer shoulders of peak--2 and peak--3 are a
result of two induced anti-crossing gaps between coupled 2DEG-like
and graphene-like plasmons. If the 2DEG in an InAs QW is removed
from our system (no QW, black dash-dotted curve), we observe that
the SP peak is strengthened in comparison with the full system (red
solid curve) and shifted towards the left (indicated by two
solid-line arrows) due to turning-off the 2DEG plasmon and its
coupling to the SP in the system. Furthermore, the graphene-like
plasmon peaks superposed on the shoulder of the SP peak become
greatly weakened due to the loss of the strong coupling between the
2DEG and graphene plasmons. Here, a simple Drude-type
optical-conductivity model\,\cite{ziegler,soljacic} cannot be
applied to our system due to the large wavenumber involved in the
deep sub-wavelength regime for a higher-order diffracted EM field.
When only the GMRA exists in the system (ribbon only, green dashed
curve), a downward-shifted dominant plasmon peak at
$\hbar\omega=14.0$\,meV, along with several kinks (a shoulder) below
(above) the peak, show up in this panel, as explained in
Fig.\,\ref{f2}. From this figure we know that the graphene-ribbon
effect can be best observed from hybridized GMRA-QW plasmon modes in
the absence of the SP. In the lower panel of Fig.\,\ref{f6}, we
display the calculated transmitted $p$--polarized E-field intensity
$\left.|{\bf E}_>(x,z\right|\omega)|^2$ at $\hbar\omega=10.7$\,meV
for the full system. In this case, a resonant SP effect on the
intensity distribution is expected around the QW-mediated air/GaAs
interface. Indeed, the intensity in the gap between two neighboring
graphene micro-ribbons, where the QW-mediated air/GaAs interface is
hosted, decreases greatly due to strong optical absorption by the SP
($\hbar\gamma_0=5$\,meV). At the same time, the intensity is built
up at two edges of a micro-ribbon and spreads out to the regions
covered by micro-ribbons. A slight asymmetry in the intensity
distribution with respect to the left and right edges of the gap
region is attributed to a finite incident angle ($\theta_i=30^{\rm
o}$).
\medskip

\begin{figure}[htbp]
%\centerline{\epsfig{file=figure7.eps,width=6.0in}}
\centerline{\includegraphics[width=6.0in]{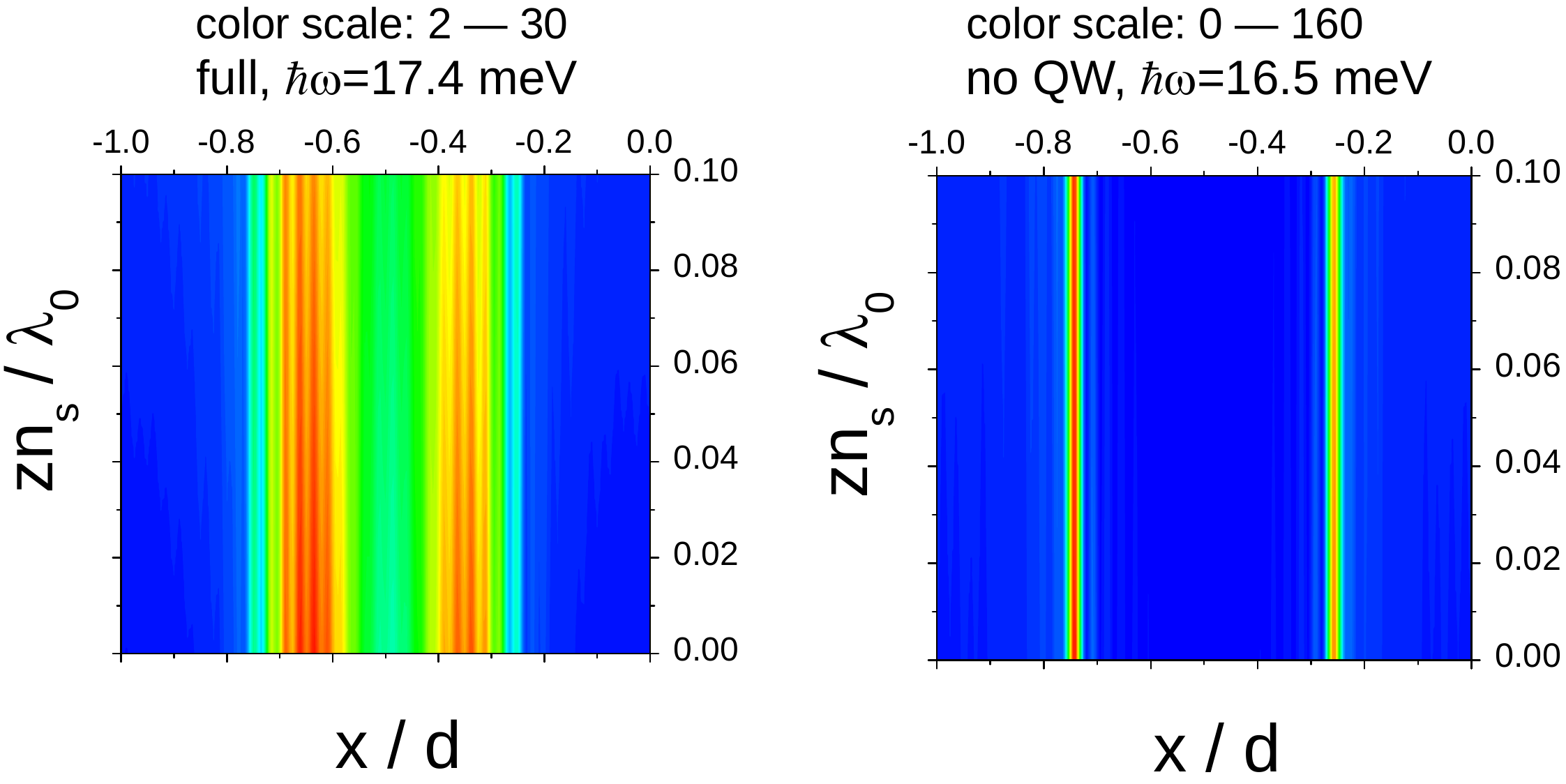}}
\caption{(Color online) A comparison of the transmitted
$p$--polarized E-field intensities $\left.|{\bf
E}_>(x,z\right|\omega)|^2$ in the presence of an SP with (left) or
without (right) a QW at two indicated resonant photon energies,
where two color scales are given in the left and right panels,
respectively. Other parameters in calculations are given in the
text.} \label{f7}
\end{figure}

In the presence of an SP, we compare the transmitted $p$--polarized
E-field intensities $\left.|{\bf E}_>(x,z\right|\omega)|^2$ in
Fig.\,\ref{f7} for the cases of with a QW at $\hbar\omega=17.4$\,meV
(left panel) or without a QW  at $\hbar\omega=16.5$\,meV (right
panel). From the right panel of this figure, we find in the
near-field regime that a GMRA tends to build up very strong E-field
intensities just at two edges of a micro-ribbon. On the other hand,
the QW would like to spread the E-field intensity across the gap
region between two neighboring micro-ribbons although two
close-magnitude maxima in intensity distribution can still be seen
between the gap center and edges. The competition of these two
opposite effects constitutes a strong EM coupling between the 2DEG
and graphene plasmon excitations, as shown in the upper panel of
Fig.\,\ref{f6}. In addition, we notice that the E-field intensity
for both panels is a little higher at the left edge of the gap
region for the lower of the two graphene-like plasmon peaks in
Fig.\,\ref{f6}.
\medskip

\begin{figure}[htbp]
%\centerline{\epsfig{file=figure8a.eps,width=3.0in}
%\epsfig{file=figure8b.eps,width=3.0in}}
\centerline{\includegraphics[width=3.0in]{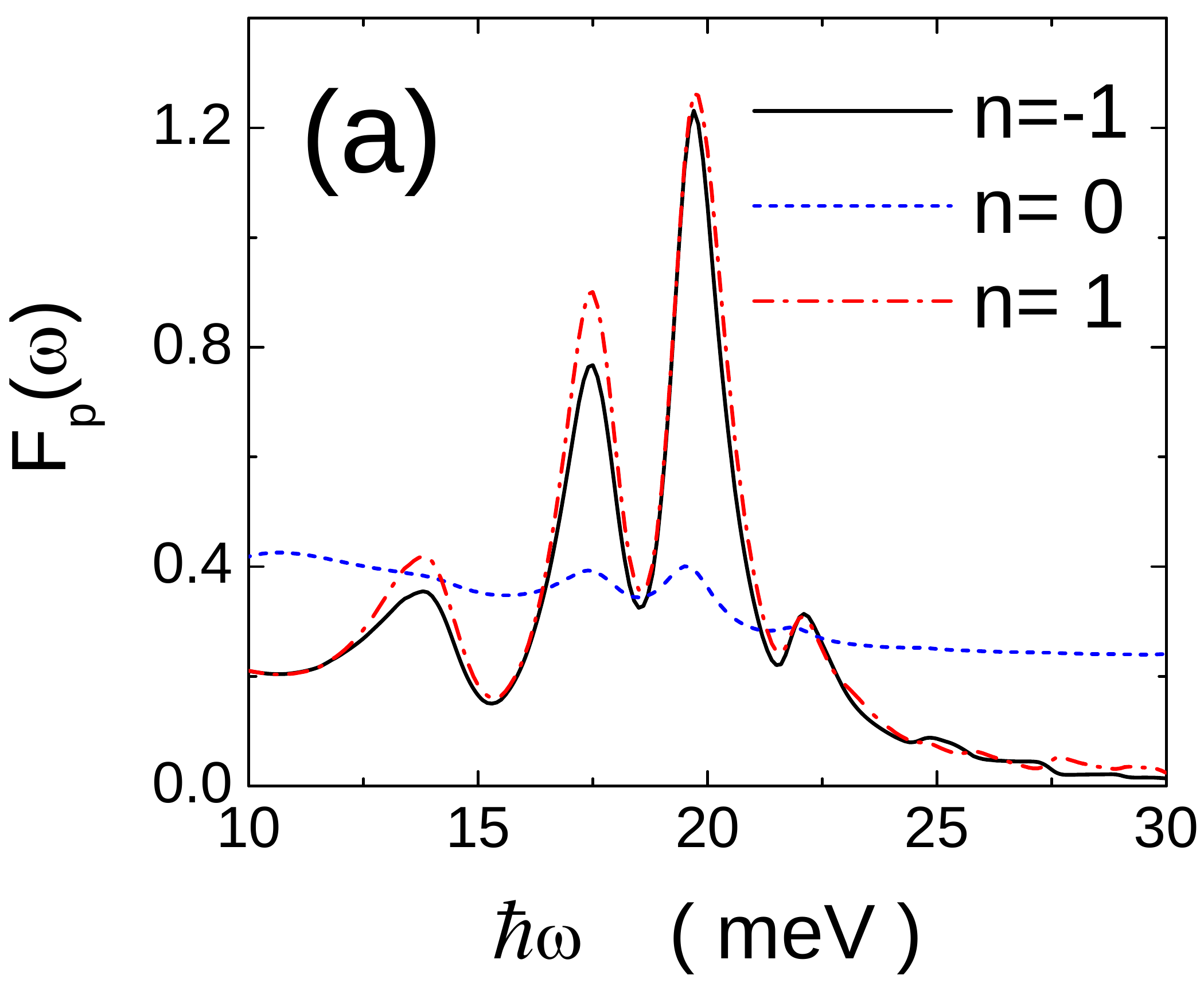}}
\centerline{\includegraphics[width=3.0in]{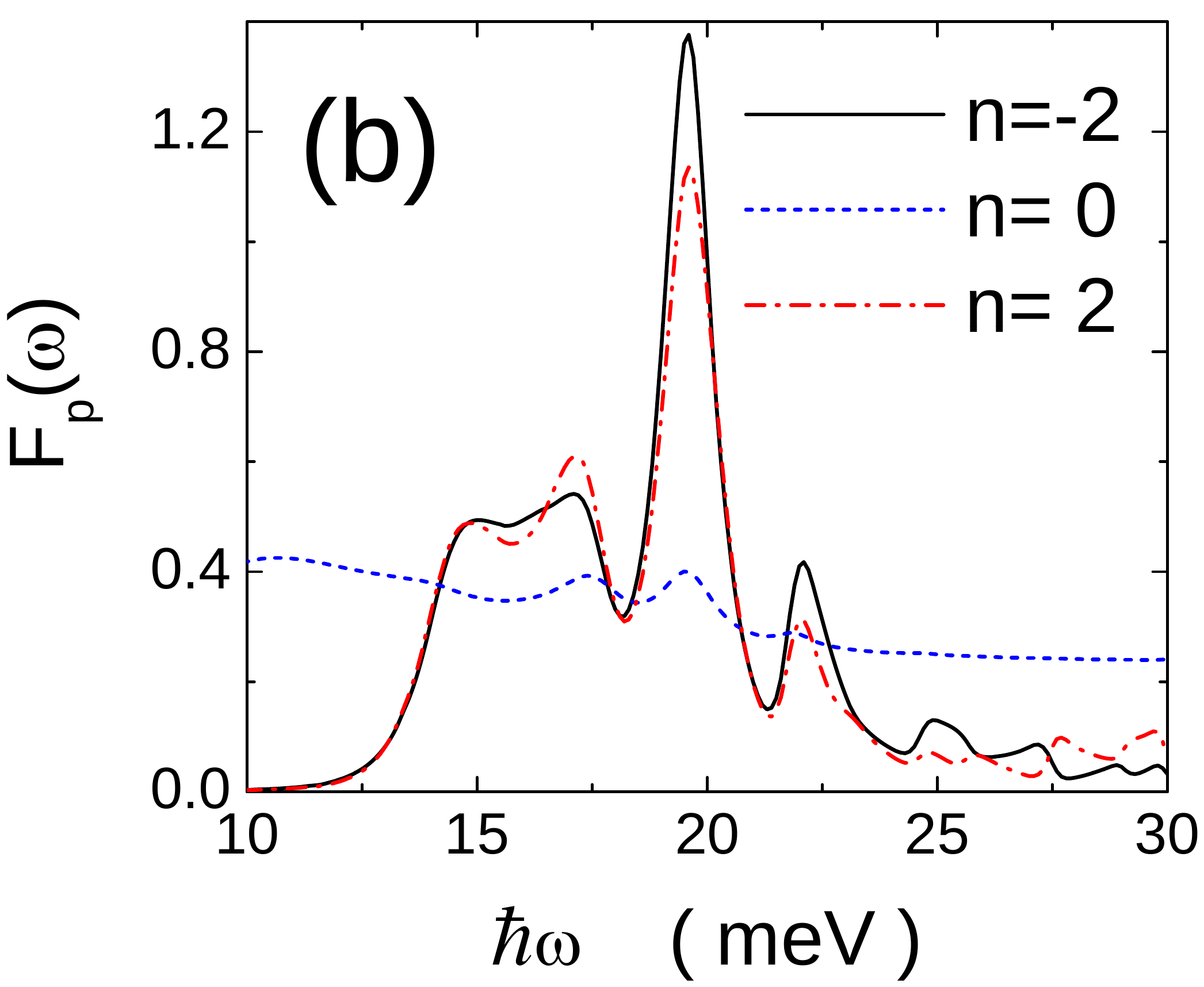}}
\caption{(Color online) Comparisons of the calculated partial near-field transmissivity
spectra $\left.F_n(q_n\right|\omega)$ for $p$ polarization. In (a), we take $n=-1$ (black solid curve),
$n=0$ (blue dashed curve) and $n=1$ (red dash-dotted curve), while in (b) we choose
$n=-2$ (black solid curve),
$n=0$ (blue dashed curve) and $n=2$ (red dash-dotted curve).
Other parameters in calculations are given in the text.} \label{f8}
\end{figure}

To find the strongly-diffracted near-field effect in the deep
sub-wavelength regime, as shown in Fig.\,\ref{f7}, on the far-field
transmissivity spectrum, we compare in Fig.\,\ref{f8} the calculated
partial near-field transmissivity spectra
$\left.F_n(q_n\right|\omega)$ for $p$ polarization, defined by
Eq.\,(\ref{e53}), with $n=\pm 1$ [in (a)] and $n=\pm 2$ [in (b)]. It
is clear from Figs.\,\ref{f8}(a) and (b) that the graphene-like
plasmon peak at $\hbar\omega=17.4$\,meV mainly comes from the
contributions of the diffraction modes with $n=\pm 1$, while the
plasmon peak at $\hbar\omega=19.5$\,meV is produced jointly by the
diffraction modes with both $n=\pm 1$ and $n=\pm 2$. In general, the
peak in the near-field transmissivity spectrum (black solid and red
dash-dotted curves) can be much stronger than that in the far-field
spectrum (blue dashed curves) due to the very strong near-field
intensity as displayed in Fig.\,\ref{f7}. The non-locality in the
optical response of electrons, as given by Eqs.\,(\ref{e16}) and
(\ref{e18}), facilitates the mixing between the EM-field specular
and diffraction modes, which enables transferring the peak weight
from a near field to a far field. We also note that the peak
strengths for the diffraction modes associated with $\pm n$ are
close in magnitude in this figure.
\medskip

\begin{figure}[htbp]
%\centerline{\epsfig{file=figure9.eps,width=3.0in}}
\centerline{\includegraphics[width=3.0in]{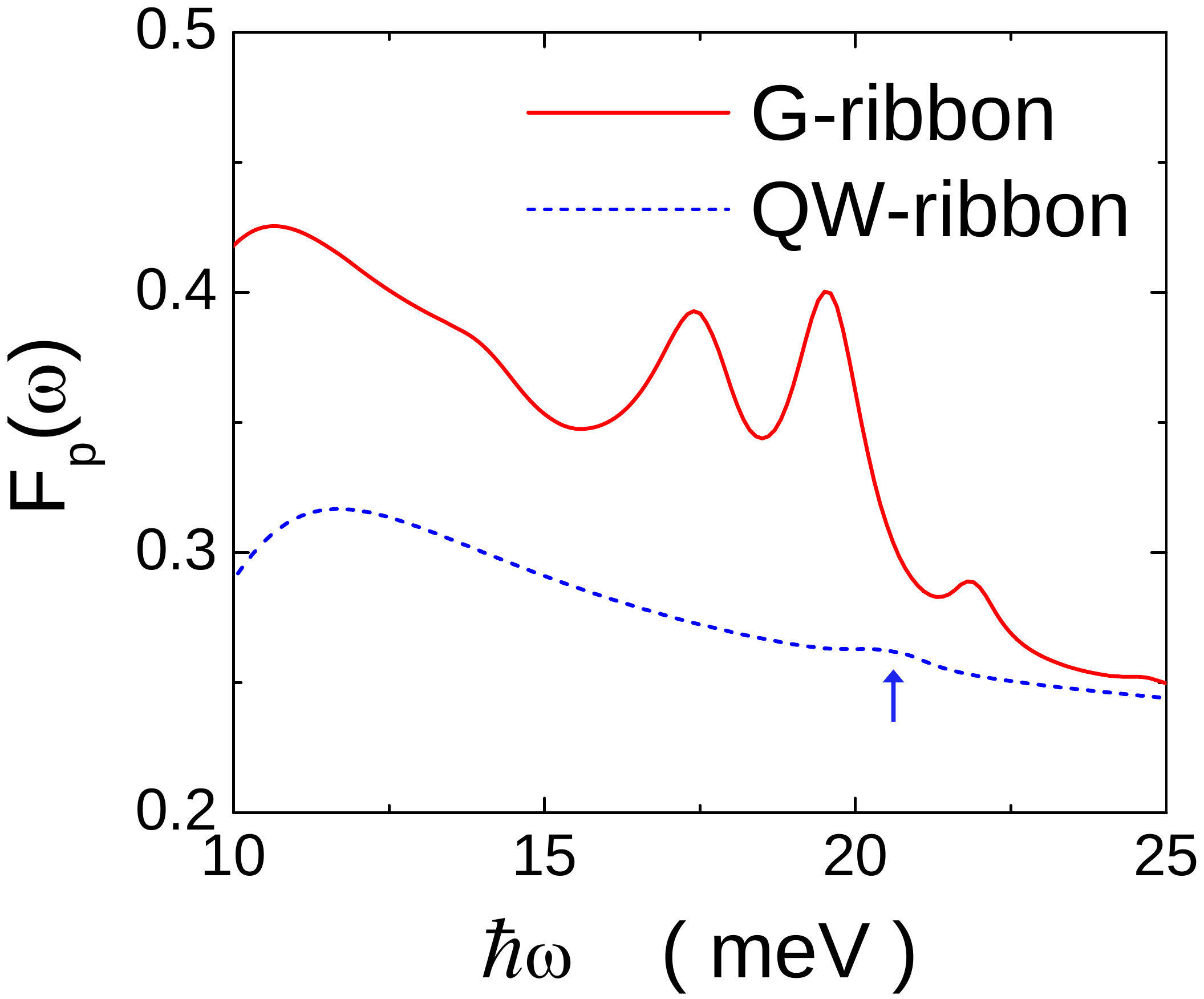}}
\caption{(Color online) A comparison of the calculated
$p$--polarized transmissivity spectra ${\cal F}_p(\omega)$ for two
compensated structures, including a graphene micro-ribbon array plus
an InAs quantum-well sheet (G-ribbon, red solid curve) and an InAs
quantum-well ribbon array plus a graphene sheet (QW-ribbon, blue
dashed curve). A blue upward arrow indicates a weak peak for the
QW-ribbon array. Other parameters in calculations are given in the
text.} \label{f9}
\end{figure}

Surprisingly, for an inverted structure (QW-ribbon, blue dashed
curve), where the graphene ribbons become a graphene sheet while the
InAs QW sheet becomes QW ribbons, we find from Fig.\,\ref{f9} that
the SP effect is greatly suppressed, compared to the original system
(G-ribbon, red solid curve), in addition to an overall reduction of
peak strength in the transmissivity. However, the EM-field
reflectivity is found to be enhanced (not shown here) in this
inverted structure for the range of photon energies shown in the
figure. In this case, only one very weak peak at
$\hbar\omega=20.1$\,meV is visible for the inverted structure, which
is attributed to an order-of-magnitude lower electron density in the
QW compared to that in a graphene micro-ribbon. At the same time,
the SP-related peak shifts from $\hbar\omega=10.7$\,meV to
$\hbar\omega=11.7$\,meV due to switching from the QW-SP coupling to
the graphene-sheet-SP coupling.
\medskip

\begin{figure}[htbp]
%\centerline{\epsfig{file=figure10a.eps,width=3.0in}
%\epsfig{file=figure10b.eps,width=6.0in}}
\centerline{\includegraphics[width=3.0in]{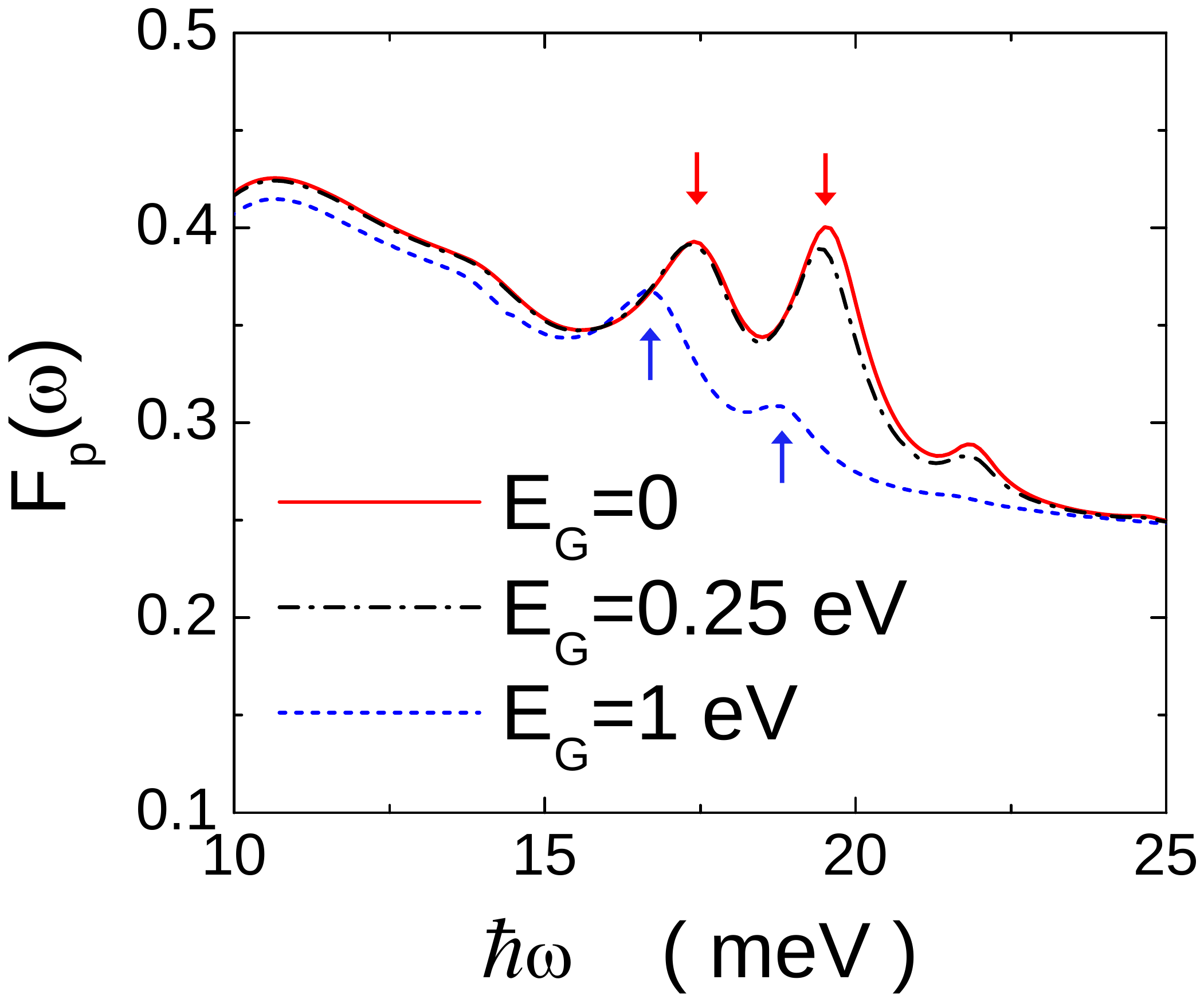}}
\centerline{\includegraphics[width=6.0in]{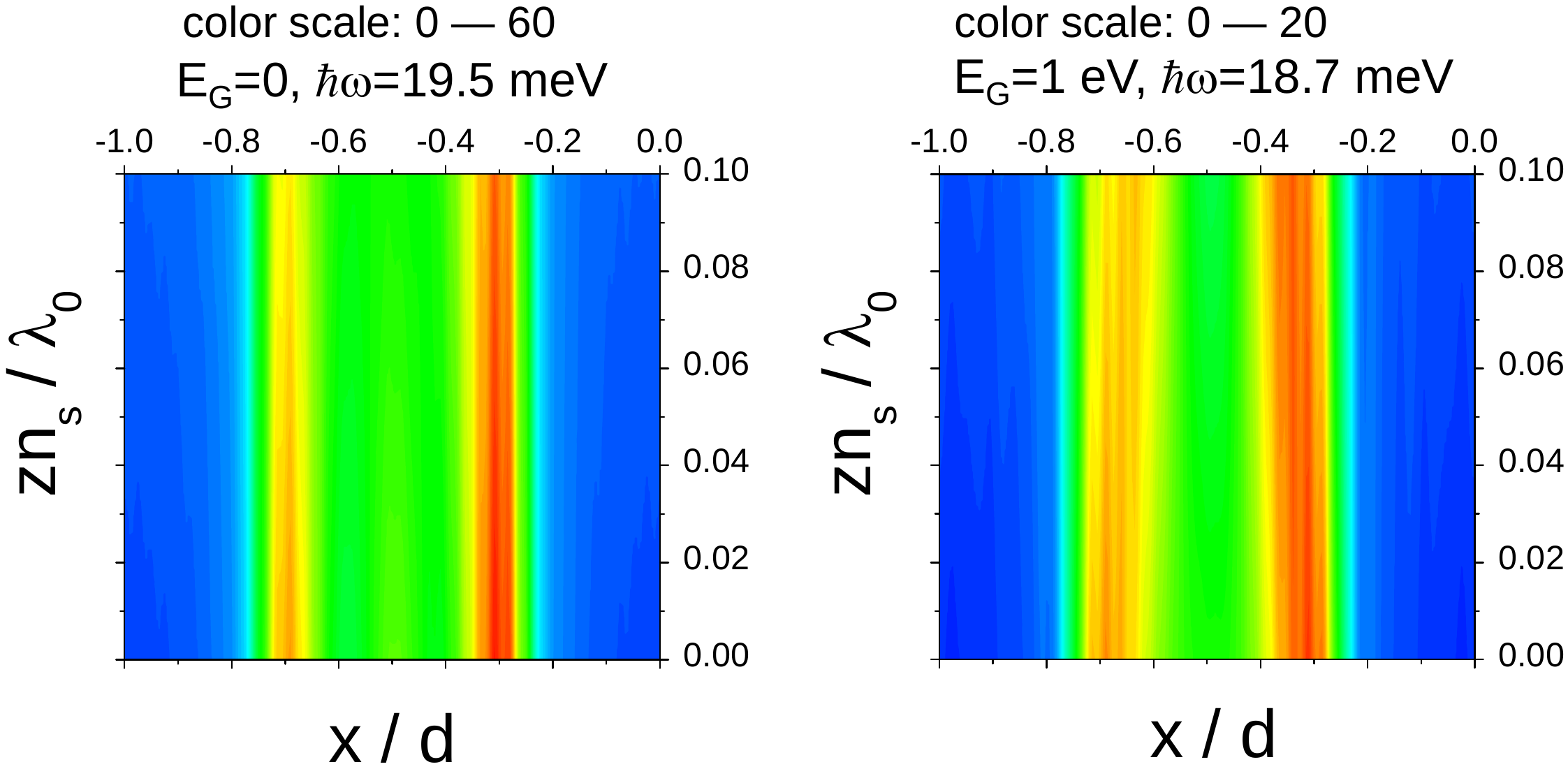}}
\caption{(Color online) A comparison of the transmissivity spectra
${\cal F}_p(\omega)$ is made in the upper panel for $p$ polarization
with three different bandgaps $E_G=0$ ($\mu_2=450$\,meV, red solid
curve), $E_G=0.25$\,eV ($\mu_2=342$\,meV, black dash-dotted curve)
and $E_G=1$\,eV ($\mu_2=173$\,meV, blue dashed curve). Two pairs of
arrows indicate the shift of a pair of corresponding peaks with
$E_G$. In the lower panel, a comparison of the calculated
transmitted $p$--polarized E-field intensities $\left.|{\bf
E}_>(x,z\right|\omega)|^2$ is displayed for $E_G=0$ (left) and
$E_G=1$\,eV (right) at $\hbar\omega=19.5$\,meV and
$\hbar\omega=18.7$\,meV, respectively. Here, we keep the electron
areal density ($\sim k_{2F}^2$) in a graphene micro-ribbon unchanged
for different values of $E_G$, and two color scales are indicated in
the lower-left and lower-right panels. Other parameters in
calculations are given in the text.} \label{f10}
\end{figure}

We display in the upper panel of Fig.\,\ref{f10} the effect of
bandgap $E_G$ on ${\cal F}_p(\omega)$ with $E_G=0$ (red solid
curve), $E_G=0.25$\,eV (black dash-dotted curve) and $E_G=1$\,eV
(blue dashed curve). Since a finite $E_G$ leads to an effective mass
($E_G/2v_F^2$) for electrons close to the band edge, its effect on
${\cal F}_p(\omega)$ is expected to be substantial (see
Fig.\,\ref{f9}) as $E_G\sim 2\hbar v_Fk_{2F}$. Indeed, we find from
the upper panel that the strengths of the two graphene-like plasmon
peaks are significantly reduced when $E_G=1$\,eV (for $\hbar
v_Fk_{2F}=0.45$\,eV). Correspondingly, from the lower panel of this
figure, we see a decrease of near-field intensity as well as an
enhanced EM coupling between the 2DEG and graphene plasmons, i.e.
the E-field intensity spreads inward from the two edges to the
center of the gap region.

\section{Conclusions}
\label{sect4}

In conclusion, we have derived in this paper a self-consistent
theory involving Maxwell equations to determine a total EM field and
the density-matrix linear-response theory to determine an induced
optical-polarization field. We have applied this self-consistent
theory to study a model system composed of an
electronically-isolated two-dimensional electron gas and an
electron-doped graphene micro-ribbon array on the surface of an
$n$-doped semi-infinite semiconductor substrate. The numerical
results for the transmissivity and reflectivity spectra in the
presence of an incident plane wave have been compared and physically
explained in the deep sub-wavelength regime for various linear-array
period, micro-ribbon width, doped graphene chemical potential and
different configurations of our model system. Our calculations have
demonstrated the existence of a tunable EM coupling among the
localized surface, conventional electron-gas and massless graphene
intraband collective excitations in the deep sub-wavelength regimes
for terahertz frequencies.

\clearpage
\noindent
{\bf Acknowledgments}\\

This research was supported by the Air Force Office of Scientific
Research (AFOSR).
\\
\medskip

\noindent
{\bf Appendix A}\\

Combining Eqs\,(\ref{e31})--(\ref{e38}), one can rewrite them into a compact
matrix equation as follow:

\begin{equation}
\tensor{M}\otimes{\bf u}={\bf b}\ , \label{a1}
\end{equation}
where the source column vector ${\bf b}$ is found to be

\begin{equation}
{\bf b}=\left[\begin{array}{c}
\left.-A_x(q_x\right|\omega)\\
\left.-A_y(q_x\right|\omega)\\
\left[(\omega/c)\,\bar{\chi}_1(q_x,\,\omega)+(\omega/c)\,\zeta\,\bar{\chi}_2(q_x,\,\omega)-(ic\eta/\omega)\right]\left.A_y(q_x\right|\omega)\\
\left[(\omega/c)\,\bar{\chi}_1(q_x,\,\omega)+(\omega/c)\,\zeta\,\bar{\chi}_2(q_x,\,\omega)-(i\omega/c\eta)\right]\left.A_x(q_x\right|\omega)\\
0\\
\vdots\\
0
\end{array}\right]\ .
\label{a2}
\end{equation}
If we limit $n=0,\,\pm 1,\,\pm 2,\,\cdots,\,\pm N$, we can write down the unknown column vector ${\bf u}$ as

\begin{equation}
{\bf u}=\left[\begin{array}{c}
\left.B_x(q_{-N}\right|\omega)\\ \vdots\\ \left.B_x(q_{N}\right|\omega)\\
\left.B_y(q_{-N}\right|\omega)\\ \vdots\\ \left.B_y(q_{N}\right|\omega)\\
\left.C_x(q_{-N}\right|\omega)\\ \vdots\\ \left.C_x(q_{N}\right|\omega)\\
\left.C_y(q_{-N}\right|\omega)\\ \vdots\\ \left.C_y(q_{N}\right|\omega)
\end{array}\right]\ .
\label{a3}
\end{equation}
Moreover, elements $M(j,\,j^\prime)$ of the coefficient matrix
$\tensor{M}$ introduced in Eq.\,(\ref{a1}) can be written out
explicitly as

\begin{eqnarray}
M(1,\,j^\prime)=\left\{\begin{array}{cc}
1 & \mbox{for $j^\prime=N+1$}
\nonumber\\
-1 & \mbox{for $j^\prime=5N+3$}
\nonumber\\
0 & \mbox{for all other $j^\prime$}
\end{array}\right.\ ,
\end{eqnarray}

\begin{eqnarray}
M(2,\,j^\prime)=\left\{\begin{array}{cc}
1 & \mbox{for $j^\prime=3N+2$}
\nonumber\\
-1 & \mbox{for $j^\prime=7N+4$}
\nonumber\\
0 & \mbox{for all other $j^\prime$}
\end{array}\right.\ ,
\end{eqnarray}

\begin{eqnarray}
M(3,\,j^\prime)=\left\{\begin{array}{cc}
-i\left(c\beta_{1,0}/\omega\right)-\left(\omega/c\right)[\bar{\chi}_1(q_x,\,\omega)+\zeta\,\bar{\chi}_2(q_x,\,\omega)]
& \mbox{for $j^\prime=3N+2$}
\nonumber\\
-i\left(c\beta_{2,0}/\omega\right) & \mbox{for $j^\prime=7N+4$}
\nonumber\\
-\left(\omega/c\right)\zeta\,{\rm sinc}(m^\prime\pi\zeta)\,\bar{\chi}_2(q_{m^\prime},\,\omega) & \mbox{for $j^\prime\in[J_1,\,J_2]\ ,\ [J_2+2,\,J_3]$}
\nonumber\\
0 & \mbox{for all other $j^\prime$}
\end{array}\right.\ ,
\end{eqnarray}
where $J_1=2N+2$, $J_2=3N+1$, $J_3=4N+2$, $m^\prime=j^\prime-(3N+2)$, and

\begin{eqnarray}
M(4,\,j^\prime)=\left\{\begin{array}{cc}
-i\left(n^2_a\omega/c\beta_{1,0}\right)-\left(\omega/c\right)[\bar{\chi}_1(q_x,\,\omega)+\zeta\,\bar{\chi}_2(q_x,\,\omega)]
& \mbox{for $j^\prime=N+1$}
\nonumber\\
-i\left(\omega/c\beta_{2,0}\right)\epsilon_s(q_0,\beta_{2,0},\,\omega)
& \mbox{for $j^\prime=5N+3$}
\nonumber\\
-\left(\omega/c\right)\,\zeta\,{\rm sinc}(m^\prime\pi\zeta)\,\bar{\chi}_2(q_{m^\prime},\,\omega) & \mbox{for $j^\prime\in[1,\,N]\ ,\ [J_1,\,J_2]$}
\nonumber\\
0 & \mbox{for all other $j^\prime$}
\end{array}\right.\ ,
\end{eqnarray}
where $J_1=N+2$, $J_2=2N+1$ and $m^\prime=j^\prime-(N+1)$.
\medskip

For $5\leq j\leq 2N+4$, we have

\begin{eqnarray}
M(j,\,j^\prime)=\left\{\begin{array}{cc}
1\ \ \ \ & \mbox{for $j^\prime=j-4\in[1,\,N]$ or $j^\prime=j-3\in[N+2,2N+1]$}
\nonumber\\
-1\ \ \ \ & \mbox{for $j^\prime=J_1\in[4N+3,\,5N+2]$ or $j^\prime=J_1+1\in[5N+4,6N+3]$}
\nonumber\\
0\ \ \ \ & \mbox{for all other $j^\prime$}
\end{array}\right.
\end{eqnarray}
with $J_1=j+4N-2$, while for $2N+5\leq j\leq 4N+4$, we acquire

\begin{eqnarray}
M(j,\,j^\prime)=\left\{\begin{array}{cc}
1\ \ \ \ & \mbox{for $j^\prime=j-3\in[2N+2,\,3N+1]$ or $j^\prime=j-2\in[3N+3,4N+2]$}
\nonumber\\
-1\ \ \ \ & \mbox{for $j^\prime=J_1\in[6N+4,\,7N+3]$ or $j^\prime=J_1+1\in[7N+5,8N+4]$}
\nonumber\\
0\ \ \ \ & \mbox{for all other $j^\prime$}
\end{array}\right.
\end{eqnarray}
with $J_1=j+4N-1$.
\medskip

In addition, for $4N+5\leq j\leq 5N+4$, this gives

\begin{eqnarray}
M(j,\,j^\prime)=\left\{\begin{array}{cc}
-i\left(c\beta_{1,m}/\omega\right)-\left(\omega/c\right)[\zeta\,\bar{\chi}_2(q_m,\,\omega)+\bar{\chi}_1(q_m,\,\omega)]
& \mbox{for $j^\prime=j-(J_1+1)$}
\nonumber\\
-i\left(c\beta_{2,m}/\omega\right) & \mbox{for $j^\prime=j+(J_1-3)$}
\nonumber\\
-\left(\omega/c\right)\,\zeta\,{\rm sinc}[(m^\prime-m)\pi\zeta]\,\bar{\chi}_2(q_{m^\prime},\,\omega) & \mbox{for $j^\prime\in[J_1,\,J_2]$ but $j^\prime\neq J_3$}
\nonumber\\
0 & \mbox{for all other $j^\prime$}
\end{array}\right.\ ,
\end{eqnarray}
where $J_1=2N+2$, $J_2=4N+2$, $J_3=j-(2N+3)$, $m=j-(5N+5)$ and $m^\prime=j^\prime-(3N+2)$,
while for $5N+5\leq j\leq 6N+4$, it leads to

\begin{eqnarray}
M(j,\,j^\prime)=\left\{\begin{array}{cc}
-i\left(c\beta_{1,m}/\omega\right)-\left(\omega/c\right)[\zeta\,\bar{\chi}_2(q_m,\,\omega)+\bar{\chi}_1(q_m,\,\omega)]
& \mbox{for $j^\prime=j-J_1$}
\nonumber\\
-i\left(c\beta_{2,m}/\omega\right) & \mbox{for $j^\prime=j+J_1-2$}
\nonumber\\
-\left(\omega/c\right)\,\zeta\,{\rm sinc}[(m^\prime-m)\pi\zeta]\,\bar{\chi}_2(q_{m^\prime},\,\omega) & \mbox{for $j^\prime\in[J_1,\,J_2]$ but $j^\prime\neq J_3$}
\nonumber\\
0 & \mbox{for all other $j^\prime$}
\end{array}\right.\ ,
\end{eqnarray}
where $J_1=2N+2$, $J_2=4N+2$, $J_3=j-(2N+2)$, $m=j-(5N+4)$ and $m^\prime=j^\prime-(3N+2)$.
\medskip

Finally, for $6N+5\leq j\leq 7N+4$, this yields

\begin{eqnarray}
M(j,\,j^\prime)=\left\{\begin{array}{cc}
-i\left(n_a^2\omega/c\beta_{1,m}\right)-\left(\omega/c\right)[\zeta\,\bar{\chi}_2(q_m,\,\omega)+\bar{\chi}_1(q_m,\,\omega)]
& \mbox{for $j^\prime=j-J_2$}
\nonumber\\
-i\left(\omega/c\beta_{2,m}\right)\epsilon_s(q_m,\beta_{2,m},\,\omega)
& \mbox{for $j^\prime=j-(J_1+1)$}
\nonumber\\
-\left(\omega/c\right)\,\zeta\,{\rm sinc}[(m^\prime-m)\pi\zeta]\,\bar{\chi}_2(q_{m^\prime},\,\omega) & \mbox{for $j^\prime\in[1,\,J_1]$ but $j^\prime\neq J_3$}
\nonumber\\
0 & \mbox{for all\ other $j^\prime$}
\end{array}\right.\ ,
\end{eqnarray}
where $J_1=2N+1$, $J_2=6N+4$, $J_3=j-(6N+4)$, $m=j-(7N+5)$ and $m^\prime=j^\prime-(N+1)$, while for $7N+5\leq j\leq 8N+4$, one is left with

\begin{eqnarray}
M(j,\,j^\prime)=\left\{\begin{array}{cc}
-i\left(n_a^2\omega/c\beta_{1,m}\right)-\left(\omega/c\right)[\zeta\,\bar{\chi}_2(q_m,\,\omega)+\bar{\chi}_1(q_m,\,\omega)]
& \mbox{for $j^\prime=j-J_2$}
\nonumber\\
-i\left(\omega/c\beta_{2,m}\right)\epsilon_s(q_m,\beta_{2,m},\,\omega)
& \mbox{for $j^\prime=j-J_1$}
\nonumber\\
-\left(\omega/c\right)\,\zeta\,{\rm sinc}[(m^\prime-m)\pi\zeta]\,\bar{\chi}_2(q_{m^\prime},\,\omega) & \mbox{for $j^\prime\in[1,\,J_1]$ but $j^\prime\neq J_3$}
\nonumber\\
0 & \mbox{for all other $j^\prime$}
\end{array}\right.\ ,
\end{eqnarray}
where $J_1=2N+1$, $J_2=6N+3$, $J_3=j-(6N+3)$, $m=j-(7N+4)$ and $m^\prime=j^\prime-(N+1)$.
\medskip

For our numerical calculations in this paper, we take
$N=40$ to ensure the accuracy of the presented results.
\\

\newpage

\begin{figure}[htbp]
%\centerline{\epsfig{file=figure6a.eps,width=3.0in}
%\epsfig{file=figure6b.eps,width=3.0in}}
\centerline{\includegraphics[width=3.0in]{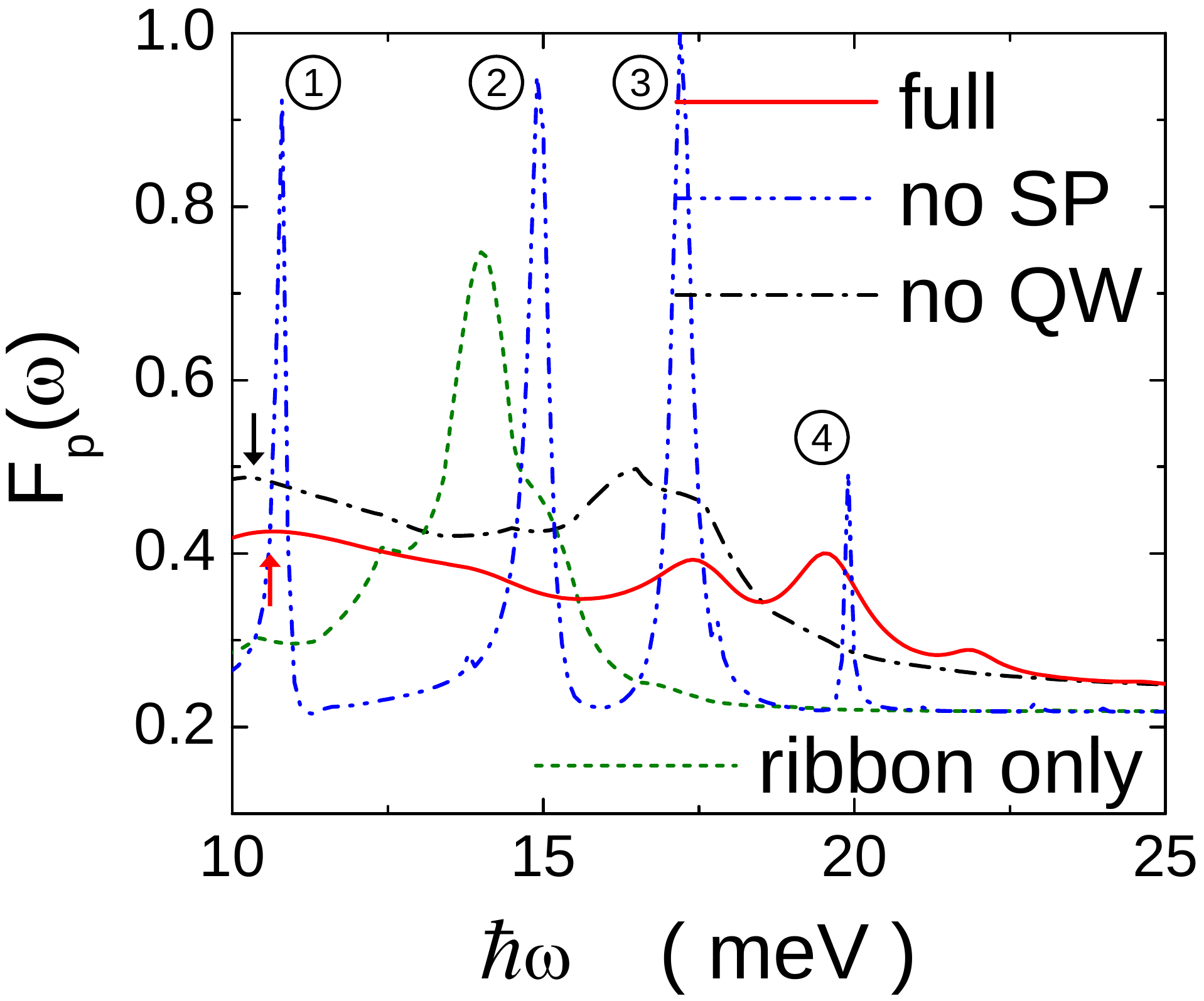}}
\centerline{\includegraphics[width=3.0in]{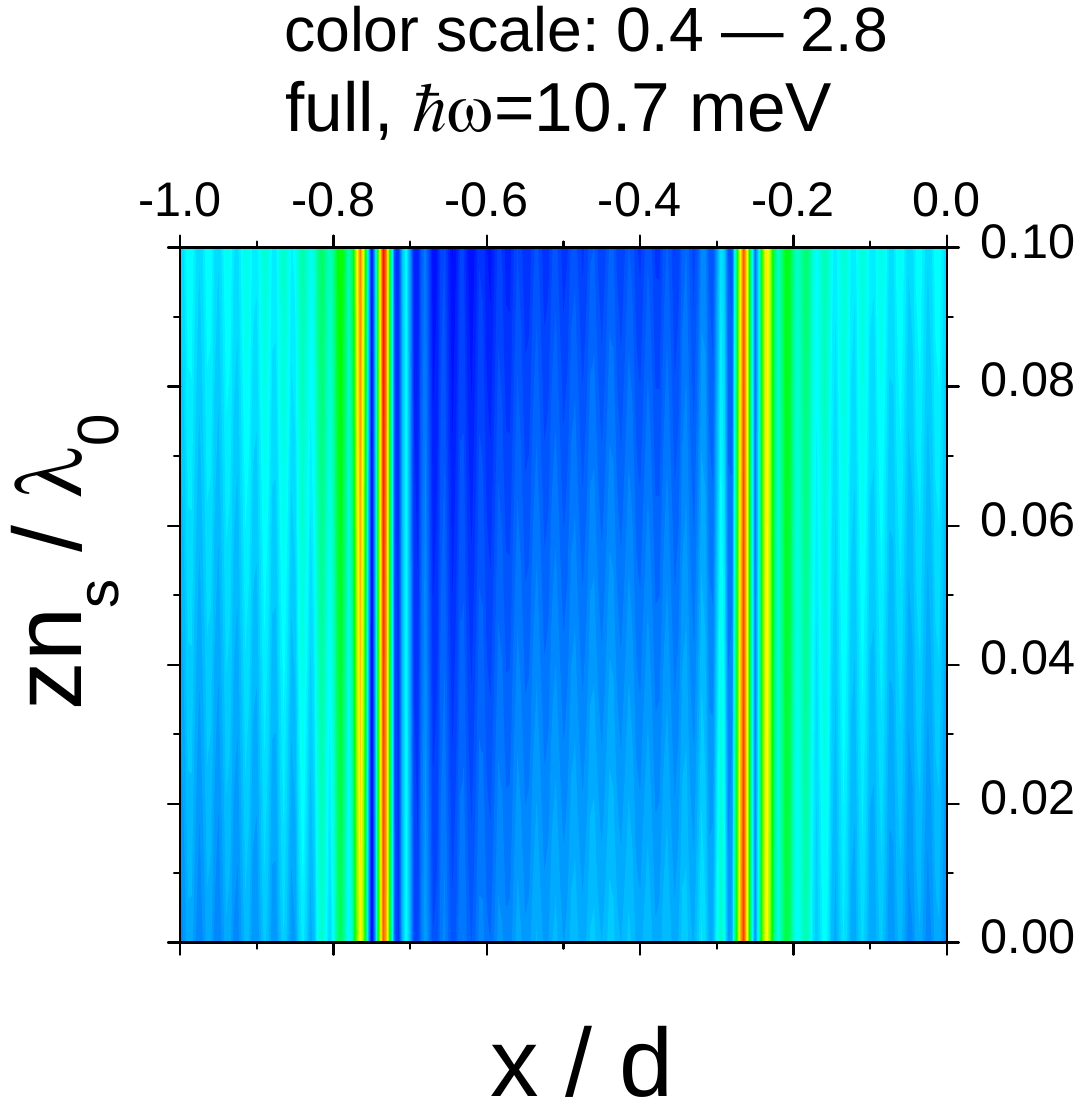}}
\caption{(Color online)) A comparison of the calculated
transmissivity spectra ${\cal F}_p(\omega)$ is presented in the upper
panel for $p$ polarization and four different configurations of the
system, including: (i) with GMRA, QW and SP (full, red solid curve);
(ii) $\Omega_{pl}=0$ (no SP, blue dash-dot-dotted curve); (iii)
$\bar{\chi}_1(q_x,\,\omega)=0$ (no QW, black dash-dotted curve);
(iv) $\Omega_{pl}=0=\bar{\chi}_1(q_x,\,\omega)=0$ (ribbon only,
green dashed curve). Two solid-line arrows indicate the peak
associated with the SP. The circled numbers label four peaks in the
figure for the case of no SP. For the lower panel, the transmitted
$p$--polarized E-field intensity $\left.|{\bf
E}_>(x,z\right|\omega)|^2$ is shown at $\hbar\omega=10.7$\,meV for
the full system, where the color scale is indicated. Other
parameters in calculations are given in the text.} \label{f6}
\end{figure}

\end{document}